\documentclass[12pt]{article}

\usepackage{amsmath}
\usepackage{amssymb}
\usepackage{mathrsfs}
\usepackage{amsthm}
\usepackage{graphicx}
\usepackage[pdftex,colorlinks=true,
urlcolor=blue,citecolor=blue,linkcolor=blue]{hyperref}

\numberwithin{equation}{section}

\newcommand{\beq}{\begin{equation}}
\newcommand{\eeq}{\end{equation}}
\newcommand{\beqa}{\begin{eqnarray}}
\newcommand{\eeqa}{\end{eqnarray}}

\def\eq#1 { \begin{equation} #1 \end{equation} }
\def\eqn#1{ \begin{eqnarray} #1 \end{eqnarray} }
\def\nn { \nonumber }
\def\C#1{\left\langle #1 \right\rangle}

\def\s{\sigma}
\def\D{\Delta}
\def\a{\alpha}

\title{On the Equivalence between Euclidean and In-In Formalisms in de~Sitter QFT}
\author{Atsushi Higuchi${}^*$, Donald Marolf${}^\dagger$, and Ian A. Morrison${}^\dagger$  \\
\\
${}^*$ Department of Mathematics, University of York\\
Heslington, York, YO10 5DD, United Kingdom \\
\href{mailto:ah28@york.ac.uk}{\texttt{ah28@york.ac.uk}} \\
\\
${}^\dagger$Physics Department, UCSB, Santa Barbara,  \\
CA 93106, USA \\
\href{mailto:marolf@physics.ucsb.edu}{\texttt{marolf@physics.ucsb.edu}}, \
\href{mailto:ian_morrison@physics.ucsb.edu}{\texttt{ian\_morrison@physics.ucsb.edu}} \\
}

\begin{document}
%%%%%%%%%%%%%%%%%%%%%%%%%%%%%%%%%%%%%%%%%%%%%%%%%%%%%%%%%%%%%%%%%

\setlength{\baselineskip}{16pt}
\begin{titlepage}
\maketitle

\begin{abstract}
We study the relation between two sets of correlators in interacting quantum field theory on de~Sitter space.  The first are correlators computed using in-in perturbation theory in the expanding cosmological patch of de~Sitter space (also known as the conformal patch, or the Poincar\'e patch), and for which the free propagators are taken to be those of the free Euclidean vacuum.
The second are correlators obtained by analytic continuation from Euclidean de~Sitter; i.e., they are correlators in the fully interacting Hartle-Hawking state.  We give an analytic argument that these correlators coincide for interacting massive scalar fields with any $m^2 > 0$.  We also verify this result via direct calculation in simple examples.  The correspondence holds diagram by diagram, and at any finite value of an appropriate Pauli-Villars regulator mass $M$.  Along the way, we note interesting connections between various prescriptions for perturbation theory in general static spacetimes with bifurcate Killing horizons.
\end{abstract}

\end{titlepage}

\tableofcontents

%%%%%%%%%%%%%%%%%%%%%%%%%%%%%%%%%%%%%%%%%%%%%%%%%%%%%%%%%%%%%%%%%
\section{Introduction}
\label{intro}
%%%%%%%%%%%%%%%%%%%%%%%%%%%%%%%%%%%%%%%%%%%%%%%%%%%%%%%%%%%%%%%%%

\begin{figure}[ht]
  \begin{center}
    \includegraphics{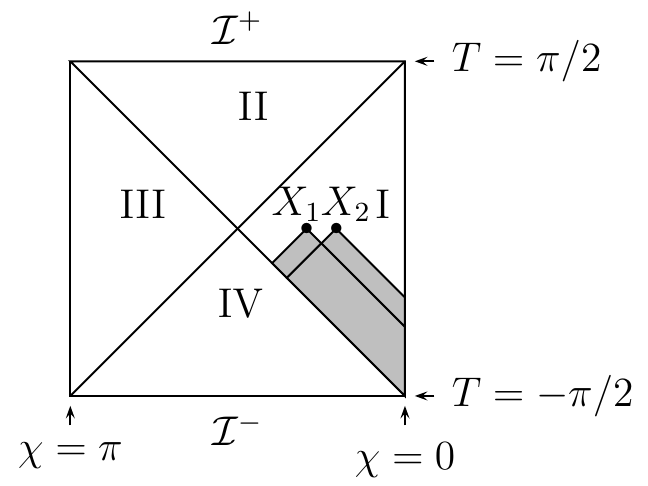}
  \end{center}
\caption{Standard Carter-Penrose diagram of de~Sitter space.  Region I is the static patch, and the Poincar\'e patch consists of regions I and II. The causal pasts of points $X_1$ and $X_2$ are the shaded regions.  See section \ref{geom} for details.}
\label{fig:one}
\end{figure}

While free quantum fields in de~Sitter space (dS${}_D$) have been well understood for some time (see \cite{Allen:1985ux} for scalar fields), interacting de~Sitter quantum field theory continues to be a topic of much discussion.  In particular, there has been significant interest in the possibility of large infrared (IR) effects in interacting de~Sitter quantum field theories
\cite{AAS,EM,Hu:1985uy,Hu:1986cv,TW,
polyakov1,PerezNadal:2008ju,Faizal:2008ns,Akhmedov:2008pu,Higuchi:2009zza,Higuchi:2009ew,Akhmedov:2009ta,Polyakov:2009nq,Burgess:2010dd,Giddings:2010nc,Krotov:2010ma}), both with and without dynamical gravity.  Most of these discussions have been in Lorentzian signature,
using some form of in-in perturbation theory. (See, e.g.~\cite{Hajicek,Kay80,Jordan:1986ug,Calzetta:1986ey} for early use of in-in perturbation theory in QFT in curved space.)  A popular choice is to choose the initial surface to be a cosmological horizon, so that the perturbation theory involves integrals over the region to the future of this horizon (see figure~\ref{fig:one}).  This region of de~Sitter space is also known as the expanding cosmological patch, the conformal patch, or the Poincar\'e patch.  We will therefore refer to the associated perturbation scheme as the Poincar\'e in-in formalism, especially when the initial state is chosen to be the free Bunch-Davies (i.e., Euclidean) vacuum.

On the other hand, IR effects are often easier to control and analyze in Euclidean signature de~Sitter space, which is just the $D$-sphere $S^D$. Analytic continuation of such correlators to Lorentz signature defines the so-called Hartle-Hawking vacuum of the theory \cite{Hartle:1976tp}.  The fact that $S^D$ is compact means that no IR divergences can arise in perturbation theory unless they are already present at order zero.  With appropriate techniques one can often analytically continue the resulting IR-finite Euclidean correlators to Lorentzian signature while maintaining control over the IR behavior.  This was done in
\cite{Marolf:2010zp,Marolf:2010nz,Hollands:2010pr} for massive scalar fields using standard perturbation theory.
For massless scalars, \cite{Rajaraman:2010xd} used the Euclidean setting to introduce a new form of perturbation theory which again yields IR-finite Euclidean correlators whose continuation to Lorentz signature can be controlled.

One would therefore like to understand precisely how correlators analytically continued from Euclidean signature are related to those computed using an intrinsically Lorentz-signature technique.  On general grounds, the analytically continued correlators will satisfy the Lorentz-signature Schwinger-Dyson equations.  So long as they satisfy appropriate positivity requirements to define a positive-definite Hilbert space, this means that the analytically-continued (Hartle-Hawking) correlators define a valid state of the theory.
Recall that positivity will generally follow from the de~Sitter analogue \cite{Schlingemann:1999mk} of reflection-positivity and the Osterwalder-Schr\"ader construction, and that reflection positivity holds formally when the Euclidean action is bounded below\footnote{This has been rigorously shown in $D=2$ dimensions for standard kinetic terms and polynomial potentials; see e.g., \cite{GJ}.}.  In such cases, it remains only to ask how the Hartle-Hawking  state relates to other states of interest, such as the state defined by in-in perturbation theory in the Poincar\'e patch.

A hint was given by \cite{Higuchi:2009ew} which studied a free scalar field but treated the mass term as a perturbation about the conformally-coupled value.  The Euclidean and Poincar\'e in-in formalisms were found to agree, and in fact to both give the exact result once all orders in perturbation theory had been included.  (There are no UV divergences due to the fact that the theory has only quadratic terms and thus only tree diagrams.) This may at first seem surprising.  Indeed, for in-in perturbation theory
defined using a Cauchy surface at finite time as the initial surface,
a result of this form would be impossible.  Since the past light cone of any external point of a Feynman diagram is cut off by the initial surface, all integrals are over regions of finite spacetime volume.  Furthermore, the volume of any such region would shrink to zero when the external point approaches the initial surface. As a result, the in-in correlators would necessarily approach the correlators of the zeroth-order theory as all arguments approach the initial slice.  On the other hand, analytic continuation of Euclidean correlators gives a de~Sitter invariant interacting state that cannot approach the zeroth-order state on any surface, so the two formalisms could not agree.

In contrast, in the Poincar\'e in-in formalism the initial surface is a null cosmological horizon.  In particular, it has the important property that there is an infinite volume of spacetime that lies both to the future of this surface and to the past of any given point in the interior of the Poincar\'e patch\footnote{This follows immediately from the fact that the Poincar\'e patch is a homogeneous space in and of itself.  Any spacetime point in the patch can be mapped to any other using only the symmetries of the patch.}.  This means that the integrals which compute perturbative corrections to the zeroth-order correlators need not become small as the arguments of correlators approach the initial surface and no contradiction with the Euclidean formalism arises.

Indeed, symmetry arguments suggest that this correspondence holds more generally. Since both the free propagators and the Poincar\'e patch is invariant under translations, rotations, and dilations, the results of Poincar\'e in-in perturbation theory will be similarly invariant so long as all integrals converge.  But for free fields on $dS_D$ the only Hadamard state which is invariant under these symmetries is the Euclidean vacuum.  One therefore expects a similar result to hold in perturbation theory, suggesting that the Poincar\'e in-in approach generally computes correlators in the interacting Euclidean vacuum.

An independent motivation comes from the work of Gibbons and Perry~\cite{Gibbons:1976pt}, who pointed out that interacting Euclidean field theory on $S^D$ describes thermal field theory inside the cosmological horizon of de Sitter space (i.e., in the static patch) with Gibbons-Hawking temperature~\cite{Gibbons:1977mu}.  While the Euclidean formalism is commonly used to study thermal field theory, there is a Lorentzian version called the Schwinger-Keldysh formalism~\cite{Schwinger:1960qe,Keldysh:1964ud}.  This formalism agrees with what is usually called the in-in formalism in relativistic field theory if the property called \emph{factorization} is satisfied (see, e.g., \cite{Landsman:1986uw}).  The physical content of this property is that generic states thermalize if given sufficient time, so that one need not take particular care to prepare a thermal state so long as the initial state is taken to be sufficiently far in the past.  Since it is known that correlators in a wide class of states approach those of the Euclidean vacuum at late times \cite{Marolf:2010zp,Marolf:2010nz,Hollands:2010pr}, it is reasonable to conjecture that the Euclidean and in-in formalisms agree at least in the static patch of de Sitter space.

We argue below that the Euclidean and Poincar\'e in-in approaches in fact agree for general interacting scalar field theories with $m^2 > 0$.  The argument can be sketched in three steps.  Step 1 is to relate the analytic continuation of Euclidean correlators to in-in perturbation theory in the static patch of de~Sitter. This amounts to checking that conditions are right for the usual relation between Euclidean field theory and Lorentz-signature thermal field theory, i.e., factorization, to hold.  Step 2 is to note that, for position-space correlators with all arguments in the static patch, in-in perturbation theory is the same whether one thinks of it as perturbation theory in the static patch or as perturbation theory in the Poincar\'e patch.  This follows from the well-known fact that in-in perturbation theory can be expressed in terms of integrals over the region that is i) to the past of all external points of a Feynman diagram and ii) to the future of the initial surface; see figure~\ref{fig:one}. As a result, analytic continuation from the Euclidean reproduces Poincar\'e in-in calculations at least when the arguments are restricted to a single static patch. Finally, step 3 is to show that both sets of correlators are appropriately analytic, so that their extension to the full spacetime is uniquely determined by their values in the static patch.  We consider Pauli-Villars regulated correlators and show agreement at each value of the Pauli-Villars regulator masses.  It follows that the fully renormalized correlators must agree as well.

The bulk of this paper is devoted to the details of this argument and to providing some simple checks of the results.  Section \ref{prelim} quickly reviews the relevant features of de~Sitter geometry.  Section \ref{factorization} then verifies that analytic continuation of Euclidean correlators does indeed give in-in correlators in the static patch for massive scalar fields, while section \ref{analyticity} argues that the correlators are sufficiently analytic so as to be determined by their restriction to the static patch.  Since the arguments are somewhat involved, we explicitly compute some simple in-in loop diagrams in section \ref{numerics} and demonstrate agreement with Euclidean results computed in \cite{Marolf:2010zp}.   We close with some discussion in section \ref{disc}.  In an appendix we describe a more direct way for the analytic continuation of Euclidean correlators, which gives a slightly different method for demonstrating their equivalence to Poincar\'e in-in correlators.

\section{Preliminaries}
\label{prelim}

This section serves to briefly review various features of both Lorentzian and Euclidean de Sitter space, and to introduce notation and conventions.  After discussing geometry and the relevant coordinate systems in section \ref{geom} we review aspects of de Sitter propagators in section \ref{prop}.

\subsection{De Sitter Geometry and Coordinates}
\label{geom}

Let us begin with Euclidean de Sitter space.  As is well known, this is just the sphere $S^D$.  Throughout this work, we set the de Sitter length $\ell$ to $1$ and work on the unit sphere.  We may thus describe $S^D$ using the metric
\begin{equation}
\label{SD}
ds_{S^D}^2 =   d\Omega_D^2 = d\vartheta^2 + \sin^2\vartheta d\Omega_{D-1}^2  , \ \ \vartheta \in [0,\pi],
\end{equation}
where $d\Omega_d^2$ is the line element of the unit $S^d$.

It is useful to consider the complexified manifold ${\mathbb S}^D$, which may be thought of as the surface $X\cdot X = 1$  in ${\mathbb C}^{D+1}$.  Wick rotations of various coordinates correspond to passing from one real section of ${\mathbb S}^D$ to another, e.g. from $S^D$ to $dS_D$. One useful Wick rotation is given by defining
\begin{equation}
\label{globalt}
  \Theta = i\left(\vartheta - \frac{\pi}{2}\right)
\end{equation}
and taking $\Theta$ real; i.e., by Wick rotating the polar angle.  This yields
\begin{equation}
\label{globalg}
ds^2_{global \ dS_D} = - d\Theta^2 + \cosh^2 \Theta^2 d\Omega_{D-1}^2  , \ \ \Theta \in {\mathbb R},
\end{equation}
which is the metric of $dS_D$ in so-called global coordinates.  Indeed, these coordinates are regular on all of $dS_D$.
Making a further coordinate transformation
\begin{equation}
\tan T = \sinh \Theta
\end{equation}
and writing $d\Omega_{D-1}^2 = d\chi^2 + \sin^2\chi d\Omega_{D-2}^2$, we have
\begin{equation}
ds^2_{global\ dS_D} = \sec^2 T(-dT^2 + d\chi^2 + \sin^2\chi d\Omega_{D-2}^2) \ \ T \in (-\pi/2,\pi/2),  \label{Carter-Penrose}
\end{equation}
where the factor inside the parentheses is the metric on a piece of the Einstein Static Universe. Note that this piece extends only for a finite amount of Einstein Static Universe time.  Figure~\ref{fig:one} is the corresponding Carter-Penrose diagram.

\begin{figure}[ht]
\begin{center}
  \includegraphics{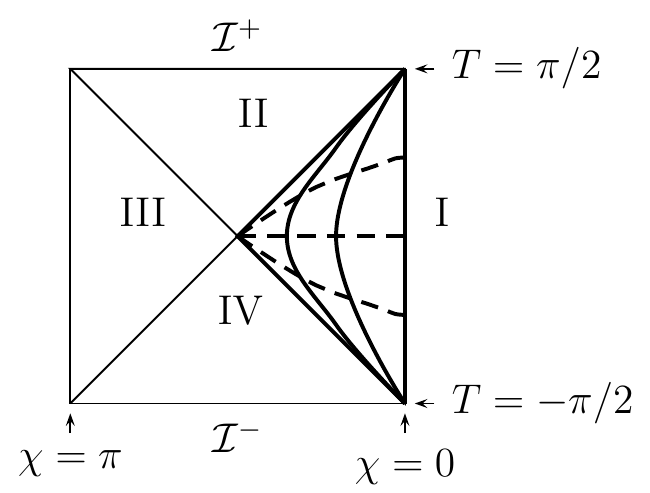}
\end{center}
\caption{Carter-Penrose diagram of de~Sitter space with $\theta = {\rm const}$ surfaces (schematically) indicated by solid lines and $t={\rm const}$ surfaces by dashed lines.}
\label{fig:two}
\end{figure}

However, one may also arrive at the same real section $dS_D$ by defining
\begin{equation}
\label{statict}
  t = i \phi, \ \ \ {\rm for} \ \ \ \tan \phi = \frac{X^1}{X^2},
\end{equation}
where $X=(X^1,X^2,\ldots,X^{D+1})$,
and taking $t$ real; i.e., by Wick rotating the azimuthal angle.  This yields
\begin{equation}
\label{staticg}
ds^2_{static \ dS_D} = - \cos^2 \theta dt^2 + d\theta^2 + \sin^2 \theta d\Omega_{D-2}^2  , \ \ t \in {\mathbb R}, \ \theta \in [0, \pi/2),
\end{equation}
with
\begin{equation}
\tan\theta = \sqrt{\frac{(X^3)^2+\cdots+(X^{D+1})^2}{(X^1)^2+(X^2)^2}},  \label{theta-def}
\end{equation}
which is the metric of $dS_D$ in so-called static coordinates.  The coordinate range $t \in {\mathbb R},  \theta \in [0, \pi/2)$ describes the static patch of de Sitter.  The coordinates $t$ and $\theta$ can be expressed in terms of $T$ and $\chi$ as
\begin{eqnarray}
\tanh t & = & \sin T\sec \chi,\\
\sin\theta & = & \sec T\sin\chi.
\end{eqnarray}
The boundary at $\theta =\pi/2$ is a coordinate singularity that coincides with the past and future cosmological horizons, $T=\pm(\chi-\frac{\pi}{2})$, defined by the observer at $\theta =0$; see figure~\ref{fig:two}.

We will also make use of so-called Poincar\'e (also known as conformally flat) coordinates on $dS_D$ in which the metric takes the form
\begin{equation}
\label{Pcoords}
ds^2 = \frac{1}{\lambda^2}(-d\lambda^2 + d\mathbf{x}^2),
\end{equation}
where $\mathbf{x}=(x^1,\ldots,x^{D-1})$.
These coordinates are related to the global ones via
\begin{eqnarray}
\lambda & = & \frac{\cos T}{\sin T + \cos\chi},\\
x^i & = & \frac{\sin \chi}{\sin T + \cos\chi}\hat{X}^i,
\end{eqnarray}
where $\hat{X}^i = X^{i+2}/\sqrt{(X^3)^2+\cdots+(X^{D+1})^2}$.
The expanding cosmological patch is the region $0 < \lambda < \infty$ with ${\mathbf x} \in {\mathbb R}^{D-1}$, which we also call the conformal or Poincar\'e patch.  Here $\lambda = \infty$ is the (past) cosmological horizon defined by the observer at ${\mathbf x}=0$, which we take to coincide with the geodesic $\theta =0$.  With this convention,  the Poincar\'e patch contains the static patch as shown in figure~\ref{fig:one}. We also take $\lambda=0$ to coincide with both $t=+\infty$ and $\Theta =+\infty$ on this geodesic. (Thus, the variable $\lambda$ runs backwards in time.  It is more common to use the variable $\eta = -\lambda$ in the cosmology community.)

\begin{figure}[ht]
\begin{center}
  \includegraphics{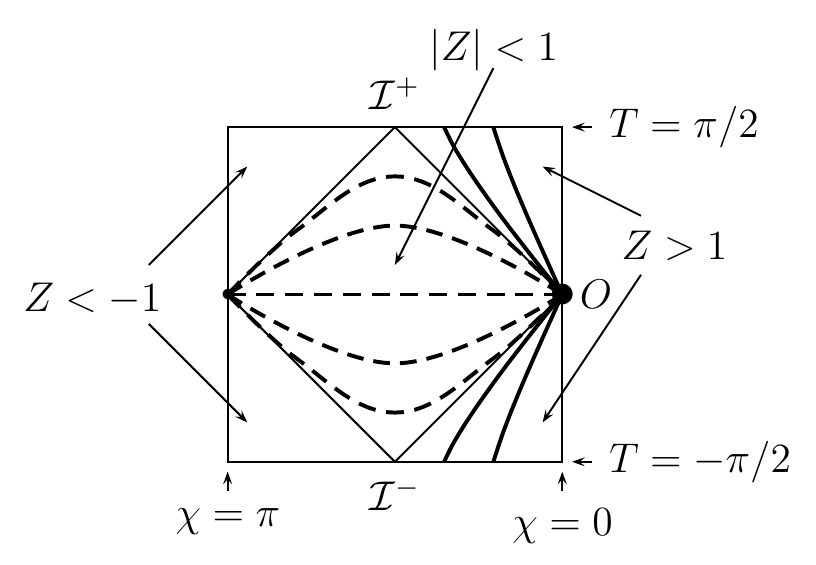}
\end{center}
\caption{Carter-Penrose diagram of de~Sitter space with timelike geodesics from point $O$ drawn with solid lines and spacelike geodesics from it drawn with dashed lines.}
\label{fig:three}
\end{figure}

The remaining relation between Poincar\'e coordinates and those discussed before is best summarized by using the concept of embedding coordinates.  Recall that $dS_D$ can be defined as the locus of points $X\cdot X = 1$ in $D+1$ dimensional Minkowski space.  Given two such points, $X$ and $Y$, one may treat them as vectors and compute the invariant Minkowski scalar product $Z = X \cdot Y$, which gives a de Sitter invariant measure of the separation between $X$ and $Y$.  In the above coordinate systems one finds
\begin{eqnarray}
 Z &=& - \sinh \Theta_x \ \sinh \Theta_y + \cosh \Theta_x \ \cosh \Theta_y \cos \gamma^{D-1}, \ \ \ {\rm (global)}\ \  \\
  &= & \cos\theta_x\cos\theta_y \cosh(t_x-t_y) + \sin\theta_x\sin\theta_y\cos\gamma^{D-2}, \ \label{invS}
  {\rm (static)}\ \ \\
\label{invP}
 &=& 1 - \frac{\|\mathbf{x}-\mathbf{y}\|^2 - (\lambda_y-\lambda_x)^2}{2\lambda_x\lambda_y}, \ \ \ (\text{Poincar\'e})
\end{eqnarray}
where $\gamma^d$ is the angle between the $X$ and $Y$ on the relevant $S^{d}$.  It is useful to note that $Z=1$ for $X=Y$ or for points connected by a null geodesic, $Z > 1$ for points connected by a timelike geodesic, $|Z| < 1$ for points connected by a spacelike geodesic, and $Z < -1$ for points which cannot be connected by any geodesic in real de Sitter space.  In the latter case, the points are not causally related; see figure~\ref{fig:three}.
Note that $Z>-1$ in the static patch.  Thus, if points $X$ and $Y$ are in the static patch, then there is a geodesic connecting these two points.

On complex de Sitter space we may take $t = \sigma + i \tau$
in static coordinates  to write
\begin{eqnarray}
Z & = &  \cos\theta_x\cos\theta_y\left[\cosh(\sigma_x-\sigma_y)\cos(\tau_x-\tau_y) - i\sinh(\sigma_x-\sigma_y)\sin(\tau_x-\tau_y)\right] \nonumber \\
& +& \sin\theta_x\sin\theta_y\cos\gamma^{D-2}, \label{invCS}
%&=& \tfrac{1}{2}\cos\theta_x\cos\theta_y \left(z_x/z_y + z_y/z_x \right) + \sin\theta_x\sin\theta_y\cos\gamma^{D-2} \ \label{invCS},
\end{eqnarray}
so that
\begin{eqnarray}
\label{magZ}
|Z|^2 & = & |\cos\theta_x\cos\theta_y\cosh(\sigma_x-\sigma_y)\cos(\tau_x-\tau_y)+ \sin\theta_x\sin\theta_y\cos\gamma^{D-2}|^2\nonumber \\
&&
+ \cos^2\theta_x\cos^2\theta_y\sinh^2(\sigma_x-\sigma_y)\sin^2(\tau_x-\tau_y),\ \ \theta_x,\theta_y \in [0,\pi/2).\nonumber \\
\end{eqnarray}

\subsection{De Sitter Propagators}
\label{prop}

Consider two points $X,Y$ on Euclidean de Sitter $S^D$.  In terms of $Z = X \cdot Y$,
the scalar propagator on $S^D$ is \cite{Bunch:1978yq,Allen:1985wd}
\begin{equation}
\label{sprop}
\Delta(X,Y) =  \frac{\Gamma(a_+)\Gamma(a_-)}{2(2\pi)^{\frac{D}{2}}\Gamma\left(\tfrac{D}{2}\right)}
(1-Z)^{\frac{2-D}{2}}F\left(\tfrac{D}{2}-a_+,\tfrac{D}{2}- a_-;\tfrac{D}{2};\tfrac{1+Z}{2}\right),
\end{equation}
where
\begin{equation}
a_{\pm}  =  \tfrac{1}{2}\left[D-1\pm\sqrt{(D-1)^2-4m^2}\right].\\
\end{equation}
Here $F$ is Gauss' hypergeometric function:
\begin{equation}
F(a,b;c;x) = 1 + \sum_{n=1}^\infty \frac{a(a+1)\cdots(a+n-1)b(b+1)\cdots(b+n-1)}{n!c(c+1)\cdots(c+n-1)}x^n.
\end{equation}

We will be interested in the analytic properties of (\ref{sprop}) for general complex $Z$.  The only singularities are branch points\footnote{These are poles if $D$ is even and if the scalar is conformally coupled and massless.} at $Z=1$ and $Z = \infty$, and we take the branch cut to connect these points along the positive real axis.  It will be particularly important to understand the singularity structure in terms of static coordinates (\ref{staticg}).  Careful inspection of (\ref{invCS}) shows the following:

\medskip

\noindent
{\bf Observation}.  The Green's function for two points $X,Y$ with static coordinates $(t_x,\theta_x)$ and $(t_y,\theta_y)$ with $\theta_x,\theta_y \in [0, \pi/2)$ is analytic for all complex $t_x,t_y$ except when $t_x-t_y$ is real modulo $2\pi i$
(so that the two points lie on the same Lorentz-signature real section)
and the two points obtained by replacing $t_x$ and $t_y$ by ${\rm Re}\ t_x$ and ${\rm Re}\ t_y$, respectively,
are causally related.  (within this real section).

\medskip

It will be useful to regulate the divergences of (\ref{sprop}) at $Z=1$ using Pauli-Villars subtractions both for the internal and external propagators so that all propagators become bounded functions of $Z$.  Because the unbounded nature of the external propagators needs to be taken into account only in the coincidence limit, where the vertex integral is convergent due to the small integration measure, it is in fact possible to show the equivalence of the Poincar\'e and Euclidean formalisms regulating only the internal propagators.  However, since analyzing such issues in detail would make the argument more cumbersome, we choose to regulate the external propagators as well.

For each $m,D$ we define a regulated propagator
\begin{equation}
\label{Dreg}
\Delta^{\rm reg}(X,Y) = \Delta(X,Y) + \sum_{i=1}^{[D/2]} C_i \Delta_{M_i}(X,Y),
\end{equation}
where $[D/2]$ denotes the integer part of $D/2$, $\Delta_{M_i}(X,Y)$ is the propagator (\ref{sprop}) for a particle of mass $M_i$, and $C_i$ are constants.  We will always assume $M_i \gg 1$ in units of the de Sitter scale, so that in particular the masses $M_i$ correspond to principal series representations \cite{Vilenkin91} of the de Sitter group.  One may choose the coefficients $C_i$ so that $\Delta^{\rm reg}(X,Y)$ has a well-defined finite limit as $Z \to 1$ (see, e.g., \cite{Camporesi:1992wn} for $D=4$). For $D =2,3$ we have $[D/2]=1$ and one may take $C_1 = -1$ for any $M_1$.  For $D=4,5$ one may choose any $C_1,C_2,M_1,M_2$ which satisfy $C_1 + C_2 = -1$ and $C_1 M_1^2 + C_2 M_2^2 = - m^2$. Nevertheless, $\Delta^{\rm reg}(X,Y)$ is not analytic at $Z=1$.  Instead, $Z=1$ remains a branch point analogous to that of the function $x \ln x$ or $x^{1/2}$ at $x=0$.

If desired, one can also make further subtractions to define regulated propagators with continuous (and thus bounded) derivatives to any specified order.  Such additional subtractions are useful in treating theories with derivative interactions, or for consideration of field-renormalization counter-terms.  Below, we will focus on non-derivative interactions for which the above subtractions will suffice.  But it will be clear from the argument that the same results hold for derivative interactions so long as an appropriate number of additional Pauli-Villars subtractions have been made.

Finally, it is useful to study $\Delta(X,Y)$ at large $|Z|$.  There, $\Delta$  behaves either like $|Z|^{-a_-}$ (for $m^2 < (D-1)^2/4$) or $|Z|^{-(D-1)/2}$ (for $m^2\geq (D-1)^2/4$).  Hence for given choices of regulator parameters $C_i,M_i$ the modulus of the regularized propagator $|\Delta^{\rm reg}(Z)|$ is bounded.   It is useful to take each $C_i,M_i$ to be a given function of the smallest regulator mass $M$, so that the regulator is removed as $M \to \infty.$  We may then take the bound on $|\Delta^{\rm reg}(Z)|$ to be $B(M)$, determined only by $m$ and the lightest regulator mass $M$.

\section{Euclidean correlators vs. thermal static patch correlators}
\label{factorization}

We now turn to step 1 of the argument sketched in the introduction.  Our task here is to show that the analytic continuation of Euclidean correlators is equivalent to those computed using in-in perturbation theory (defined using the propagator of the free Euclidean vacuum) in the so-called static patch of de Sitter. This essentially amounts to checking that conditions are right for the usual relation between Euclidean field theory and Lorentz-signature thermal
field theory to hold; i.e., that the Hartle-Hawking correlators are indeed thermal correlators in the static patch.  At a formal level, this follows from the
fact that correlation functions $Tr [\phi(x_1)...\phi(x_n) e^{-\beta H}]$ in the canonical ensemble are given by an imaginary-time path integral; see e.g., \cite{Landsman:1986uw}.  However, in order not to miss any subtleties (perhaps due to IR divergences of the sort predicted in \cite{Polyakov:2009nq}) and because of the many controversies surrounding dS quantum field theory, we will proceed slowly through an explicit perturbative argument.    Below, we consider diagrams using the Pauli-Villars regularized propagators (\ref{Dreg}) so that $|\Delta^{\rm reg}(Z)| \le B(M)$.  We restrict attention to connected diagrams since vacuum bubbles are automatically excluded both in the Euclidean and in-in formalisms.  Because the desired result is trivial for the diagram with two external points connected by a single propagator, we also exclude this diagram from our discussion.  Non-derivative interactions are assumed for simplicity, though the argument is readily extended to derivative interactions so long as additional Pauli-Villars subtractions are made as described in section \ref{prop}  above.

Recall that in static coordinates (\ref{staticg}) points of de Sitter space are labeled by a pair $(t, \hat X)$ where $\hat X$ is a point in the (open) northern hemisphere of $S^{D-1}$. We will use these coordinates for both the static patch of Lorentz-signature $dS_D$ (where $t \in {\mathbb R})$ and on Euclidean-signature de Sitter $S^D$ (where $-i t \in (-\pi,\pi)$.) We imagine that the integrals over the time coordinates $t_i$ of the internal vertices will be performed first, followed later by the integrals over $\hat X_i$.  So for the moment we consider the $\hat X_i$ to be fixed. We also assume that all internal vertices and external points correspond to distinct spatial points $\hat X$; i.e., $\hat X_i \neq \hat X_j$ for $i \neq j$.  Due to our Pauli-Villars regularization, we can always recover information at coincidence by continuity.

Let us first review the general argument relating Euclidean correlators to in-in correlators (see e.g. \cite{Landsman:1986uw}) using our de Sitter static patch notation.  In the Euclidean approach the time integrals of the internal vertices are all from $i\pi$ to $-i\pi$.  The external points are taken to lie on this contour and, at least for the moment, we take them to all lie close to (though not necessarily precisely at) $t=0$.  Since the $\hat X_i$ are distinct, it follows from the Observation of section \ref{prop} that the integrand is analytic in all time coordinates $t_i$ in a region containing the contour of integration.  Thus the contour can be deformed.  In fact, taking all internal coordinates $t_i$ to be integrated along the same contour $C$, we note that the contour can be freely deformed so long as i) it begins at some $t = t_0 +i \pi$ with $t_0$ real and ends at $t = t_0 -i \pi$, ii) the imaginary part of $t$ is strictly decreasing everywhere (so that no two points on the path have the same value of ${\rm Im} \ t$) and iii) the path continues to pass through the external points.  In particular, we are free to take the limit $t_0 \rightarrow - \infty$.

\begin{figure}[ht]
\begin{center}
  \includegraphics{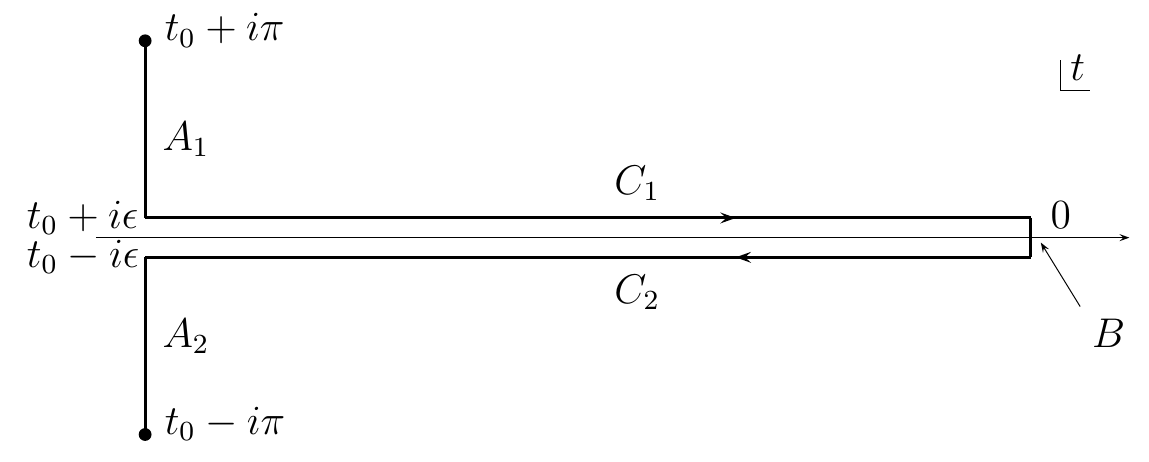}
\caption{A deformation of the Euclidean contour.}
\label{fig:four}
\end{center}
\end{figure}

Now, these rules allow us to choose the contour $C = A_1+C_1+B +C_2+A_2$ to be as in figure~\ref{fig:four}.
Here $\epsilon\, (<\pi)$ is a nonzero positive number.  The imaginary part is increasing infinitesimally on the
horizontal portions of the contour.  This is equivalent to using the Feynman propagator when the two points are both
on the upper horizontal portion and the Dyson (or anti-Feynman) propagator if the two points are both on the lower horizontal portion.  In general, in the $\epsilon \rightarrow 0$ limit (and where the imaginary parts of the times for all external points are also taken to zero), one may say that the above contour computes correlators using the free path-ordered two-point function as the propagator, just as occurs in the in-in formalism.
Furthermore, since all integrals converge after Pauli-Villars regularization, it is clear that the integral along $B$ is of order $\epsilon$ and can be neglected in the limit $\epsilon \to 0$.

As a result, the Euclidean correlators (evaluated at $t=0$) agree with the corresponding in-in correlators in the static patch (computed using the propagators of the free Euclidean vacuum) so long as a property called {\it factorization} \cite{Landsman:1986uw} holds, which states that the $A_1,A_2$ pieces of the contour $C$ can be neglected in the $t_0 \rightarrow -\infty$ limit.  We now establish this property for our systems, diagram by diagram\footnote{For a general contour $C$, we will refer to the associated diagrams below as Feynman diagrams, even though they may sometimes involve Dyson (or other) propagators as noted above.}.
For each Feynman diagram, let us choose one external point $X = (t_e, \hat X_e)$ and one internal point $Y$ that lies on either segment $A_1$ or $A_2$.  To show that the integral of $Y$ over the above segments can be neglected, we also choose a path through the diagram from $X$ to $Y$; i.e., a particular chain of propagators.

Now, recall from section \ref{prop} that at fixed Pauli-Villars regulator mass $M$ all propagators are bounded by some $B(M)$.  To establish a bound on the integrals, we may thus replace the integrand with its magnitude and replace all propagators {\it not} on the chosen path by $B(M)$.  Next consider the propagators on the chosen path.  For at least one such propagator, external or internal, the (static-patch) time coordinates of its two arguments have real parts differing by at least $({t}_e-t_0)/K$, where $K$ is the number of propagators in the chain. From (\ref{magZ}) and the asymptotics of the propagators discussed in section \ref{prop}, this means that this propagator is of order $[\cos \theta_1 \cos \theta_2 e^{({t}_e-t_0)/K}]^{-\nu}$ or smaller for some positive $\nu$ determined by the mass $m$ of the quantum fields, where $\theta_1$ and $\theta_2$ are the $\theta$-coordinates of the two arguments of this propagator, if ${t}_e - t_0$ is large enough. Replacing all other propagators on this chain with $B(M)$, we integrate the time coordinate $\tau$ of $Y$ along the segments from $t_0+i\pi$ to $t_0+i\epsilon$ and from $t_0-i\epsilon$ to $t_0-i\pi$.  We also perform all other $t$-integrals at the vertices.   The result is clearly bounded by
\begin{equation}
c_2[B(M)]^{n_1}({t}_e-t_0)^{n_2}[\cos \theta_1 \cos \theta_2 e^{({t}_e-t_0)/K}]^{-\nu}  \label{bound}
\end{equation}
for some constants  $c_2,n_1,n_2$, where the factors of
$({t}_e-t_0)^{n_2}$ come from the measure.
It is important to note that $c_2,n_1,n_2$ are independent of the positions of all vertices, as well as $t_0$.

To complete the argument, we divide the integrals over $\theta_1,\theta_2$ (or, say, just $\theta_1$ if the 2nd point is external so that $\cos \theta_2$ is fixed and independent of $t_0$) into two regions.  In the first, we take $\cos\theta_1, \cos \theta_2 > e^{-({t}_e-t_0)/3K}$. The bound in (\ref{bound}) then shows that the integral over this region tends zero at least like $({t}_e-t_0)^{n_2}e^{-2\nu({t}_e-t_0)/3K}$ as $t_0\to -\infty$.  The remaining region of integration is small since one of the variables to be integrated ($\theta_1$ and/or $\theta_2$) satisfies $\cos\theta < e^{-({t}_e-t_0)/3K}$ and the integration measure $\sin^{D-2}\theta\cos\theta\,d\theta$ contains a factor of $\cos \theta$. We note that the length of the interval on which $\theta$ is integrated is of order $e^{-(t_e-t_0)/3K}$ as well.  We may therefore replace {\it all} propagators by the bound $B(M)$ and find that the contribution from this region is again bounded by a number of the form $c_3({t}_e-t_0)^{n_3}e^{-2({t}_e-t_0)/3K}$, which of course tends to zero as $t_0\to -\infty$.  This establishes the fact that sections $A_1$ and $A_2$ can be neglected in the desired limit for any (finite or infinitesimal) choice of $\epsilon$ in figure~\ref{fig:four}.  In particular, this demonstrates the agreement of Euclidean and static patch in-in correlators (computed using the propagator of the free Euclidean vacuum) when the external points are located at $t=0$.

To demonstrate agreement for more general external points, we need only analytically continue the correlators as a function of the time coordinates of the external points.  This is in fact the {\it definition} of the Euclidean correlators evaluated at more general times, and we will show that it also gives the static patch in-in correlators.  For this step, it is convenient to take the external points to have distinct (and fixed) values of ${\rm Im} \ t$.  At the end of the argument we will take the limit where all of these imaginary parts vanish.

\begin{figure}[ht]
\begin{center}
  \includegraphics{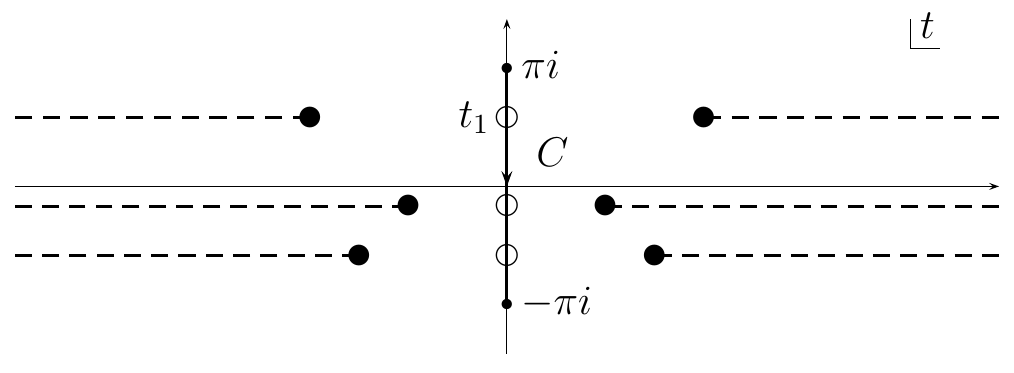}
\end{center}
\caption{Singularities and branch cuts of propagators. }
\label{fig:five}
\end{figure}

We first consider the analyticity of the integrand for some given diagram in the time coordinate $t_1$ of some external point with the time coordinates of all other points (both internal and external) held fixed and taken to lie on one of the contours $C$ discussed above.  We also take the spatial coordinates of all points to be fixed and distinct.   Due to the observation of section \ref{prop}, the singularities are then a finite distance from the contour $C$.
For example, for the original Euclidean integral, if the external point with the time coordinate $t_1$ is connected to a vertex with time coordinate $t$ which is also connected to two other vertices, the singularities and associated branch cuts on the complex $t$-plane are similar to those shown in figure~\ref{fig:five}.

\begin{figure}[ht]
\begin{center}
  \includegraphics{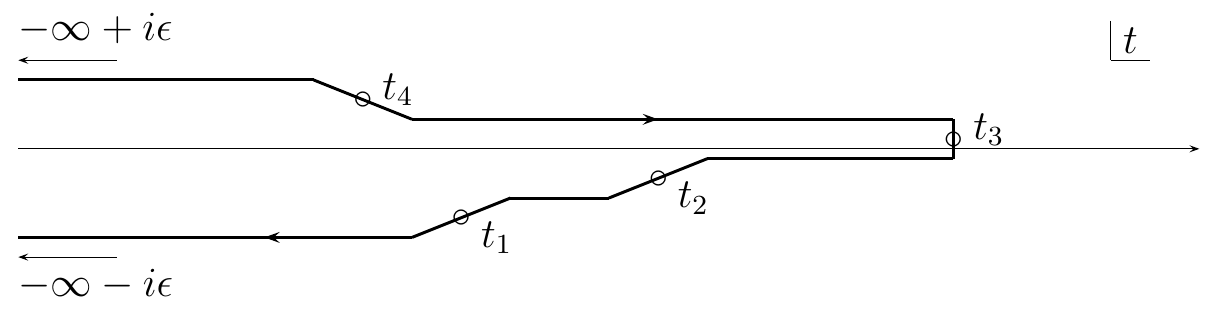}
\end{center}
\caption{Deformed contour for external points at $t_i$ with finite imaginary parts. }
\label{fig:six}
\end{figure}

\begin{figure}[ht]
\begin{center}
  \includegraphics{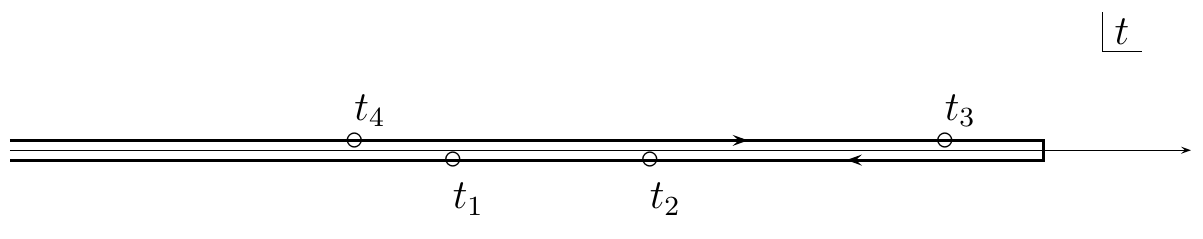}
\end{center}
\caption{$t$-integration contour for the in-in formalism in the static patch.  The open circles denote external points at times $ t_i$.}
\label{fig:seven}
\end{figure}

We may thus analytically continue $t_1$ to any complex value so long as we avoid the branch cuts.  Let us do so holding ${\rm Im} \ t_1$ fixed and distinct from the imaginary parts of all other external time coordinates.  Then the only singularities which are of concern are those due to the vertex connected to $t_1$ with the same imaginary part of $t$; i.e., for which ${\rm Im} \ t = {\rm Im} \ t_1$.  As indicated in figure~\ref{fig:five}, for a fixed contour $C$ this will in general allow only a finite range over which the integrand can be analytically continued in ${\rm Re} \ t_1$.  However, as noted earlier, we are also free to further deform the contour. For example, by shifting the contour for all vertices a bit to the right at ${\rm Im} \ t_1$, we shift the allowed window for analytic continuation a bit to the right, and we do so without changing the size of this window.  It is thus clear that, by dragging the contour along with the external point in this way, we may analytically continue the result of the time integrations  to arbitrary values of ${\rm Re} \ t_1$ for any  given distinct set of spatial coordinates.  But as before, our Pauli-Villars regularization scheme implies the same result holds for general spatial coordinates by continuity\footnote{Continuity of the integrand is clear from the regularization scheme.  Continuity of the result of the time integrations  follows from the fact that these integrals converge absolutely.  This in turn follows from the same estimates used to show factorization above.}.  It follows that the analytic continuation of Euclidean correlators can be computed via the usual Feynman diagrams associated with any contour which i) begins  at some $t = -\infty + i\epsilon$ with any real and positive $\epsilon$ ends at $t = -\infty - i\epsilon$, ii) has the imaginary part of $t$ strictly decreasing everywhere (so that no two points on the path have the same value of ${\rm Im} \ t$) and iii) passes through all external points\footnote{\label{Esreg} The reader may ask if the analytic continuation of a full diagram (after all integrals, including space integrals, have been performed) coincides with the result described above (in which the integrand is first continued, before performing the spatial integrals).  The potential obstacle is the fact that spatial coordinates will necessarily coincide somewhere during the integrals over space, and such coincidences shrink the windows (used to enact the analytic continuation above) between past- and future-branch cuts to zero size.  One may show that this is not an issue by performing a further regularization in which all  propagators $\Delta(Z)$ are replaced by $\Delta(Z-s)$ for some positive $s$.  This regularization maintains windows of finite size even at coincidence.  Furthermore, so long as one drags the contour along with the external point as described above, one finds that the resulting integral is analytic in the external time variables for all positive $s$ on the domain where the external times have distinct imaginary parts.  Then we find that  the full diagram is analytic at $s=0$, and its analytic continuation is given by the prescription above. The argument is very similar to that given in Section \ref{regA} to establish the analyticity of the Poincar\'e in-in correlators in the conformal-time variables. }.
An example is shown in figure~\ref{fig:six}.
Taking the limit $\epsilon \rightarrow 0$ (and taking the limit in which all external time coordinates now become real) gives the usual closed-time-path representation of the static patch in-in correlators (defined using the propagators of the free Euclidean vacuum) just as described above for external points at $t=0$; see figure~\ref{fig:seven}.

\section{Analyticity of in-in correlators}
\label{analyticity}

Recall that our goal is to demonstrate the equivalence of the Poincar\'e and Euclidean formalisms for perturbation theory.  We outlined a three-step argument in the introduction.  As described there, it is clear that the in-in formalism in the static patch is a restriction of that in the Poincar\'e patch (Step 2).  Since we have now shown that the static patch in-in correlators agree with those of the Euclidean formalism (Step 1), it remains only to show that Poincar\'e in-in correlators are appropriately analytic in their arguments (Step 3).  The desired result then follows since two analytic functions that agree in any non-empty open subset of a real section must in fact agree everywhere.  In this section we will establish analyticity of Poincar\'e in-in correlators as functions of the conformal-time coordinates with space coordinates fixed.  This will turn out to be sufficient for our purpose.

\begin{figure}[ht]
\begin{center}
  \includegraphics{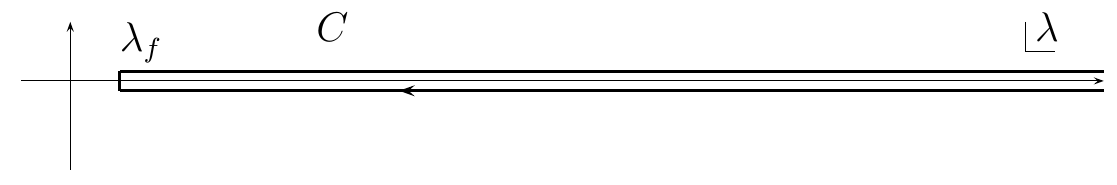}
\end{center}
\caption{Typical $\lambda$-contour for the in-in formalism in the Poincar\'{e} patch.  All external points have $\lambda > \lambda_f$}
\label{fig:eight}
\end{figure}

Recall that, for given external points $X_j=(\tilde\lambda_j,\mathbf{x}_j)$, in the coordinates of (\ref{Pcoords}) any Poincar\'e correlator is a sum of terms of the form
\begin{equation}
\mathcal{A}_{P} = c_1\left(\prod_{k=1}^n  \int d^{D-1}\mathbf{y}_k\int_C \frac{d\lambda_k}{\lambda_k^{D}}\right)
F(Y_1,\ldots,Y_n)\prod_{j=1}^{m}\Delta^{\rm reg}(X_j,Y_k),  \label{ininA}
\end{equation}
where a typical contour $C$  is shown in figure~\ref{fig:eight} and we have used the Pauli-Villars regulated propagators $\Delta^{\rm reg}(X_j,Y_k)$.    The contour is infinitesimally away from the real line, and the imaginary part of $\lambda$ \emph{increases}  infinitesimally everywhere along the contour, even on the horizontal sections.  Time-ordered correlators are obtained by putting the $\lambda$-coordinates, $\tilde\lambda_j$, of the external points on the lower horizontal line, whereas anti-time-ordered correlators are obtained by putting them on the upper horizontal line. We will refer to any such ${\mathcal A}_P$  as an amplitude, and  we will again refer to the associated diagram as a Feynman diagram even though diagrams include Dyson (or other) propagators in computing ${\mathcal A}_P$.

\begin{figure}[ht]
\begin{center}
  \includegraphics{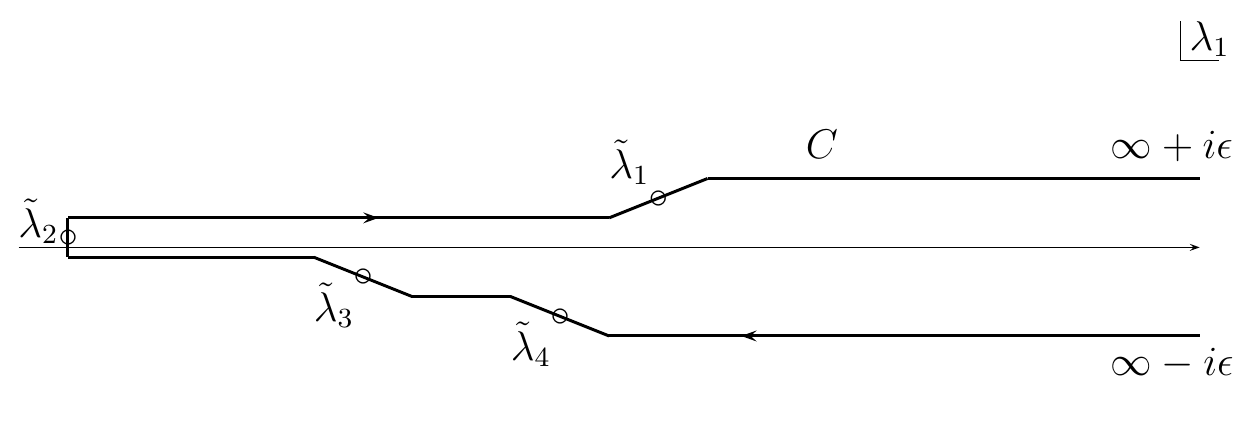}
\end{center}
\caption{Deformed contour for external points with finite imaginary parts in the Poincar\'e patch}
\label{fig:nine}
\end{figure}

Like the in-in amplitude in the static patch, the Poincar\'e in-in amplitude $\mathcal{A}_P$ is obtained by first considering the corresponding amplitude with finite and distinct imaginary parts ${\rm Im} \ \tilde \lambda_j$  of the conformal-time  coordinates of the external points and then taking the limit ${\rm Im} \ \tilde\lambda_j \to 0$.  For this reason we let $\tilde{\lambda}_j$ satisfy
${\rm Im} \ \tilde{\lambda}_1 < {\rm Im} \ \tilde{\lambda}_2 < \cdots < {\rm Im}\ \tilde\lambda_m$ without loss of generality and use the contour analogous to that considered for the static patch.  Figure~\ref{fig:nine} shows an example with $m=4$ with finite imaginary parts, before taking the $\epsilon \rightarrow 0$ limit.

It is important to note that, in general, one must integrate over the conformal-time coordinates  $\lambda_k$ first in (\ref{ininA}), before integrating over the spatial coordinates as the integrand may otherwise decay too slowly at large $\|\mathbf{y}_k\|$ for the $\mathbf{y}_k$-integrals to converge if each $\lambda_k$ is fixed on the contour.
We will show below that, with our Pauli-Villars regulators, all integrals converge so long as the $\lambda_k$-integrals are performed first.
We then use this result to demonstrate the desired analyticity of ${\mathcal A}_P$.

\subsection{Convergence of ${\mathcal A}_P$}
\label{CAP}

We now verify that integrals defining the amplitude $\mathcal{A}_P$ converge with the contour $C$ chosen as in figure~\ref{fig:nine} so long as we perform the $\lambda_k$-integrations before the $\mathbf{y}_k$-integrations.  The general strategy is  deform the $\lambda$-contour at each vertex as much to the right as possible, while avoiding singularities, so that the regions of spacetime over which the vertices are integrated become small enough to guarantee absolute convergence.

\begin{figure}[ht]
  \begin{center}
    \includegraphics{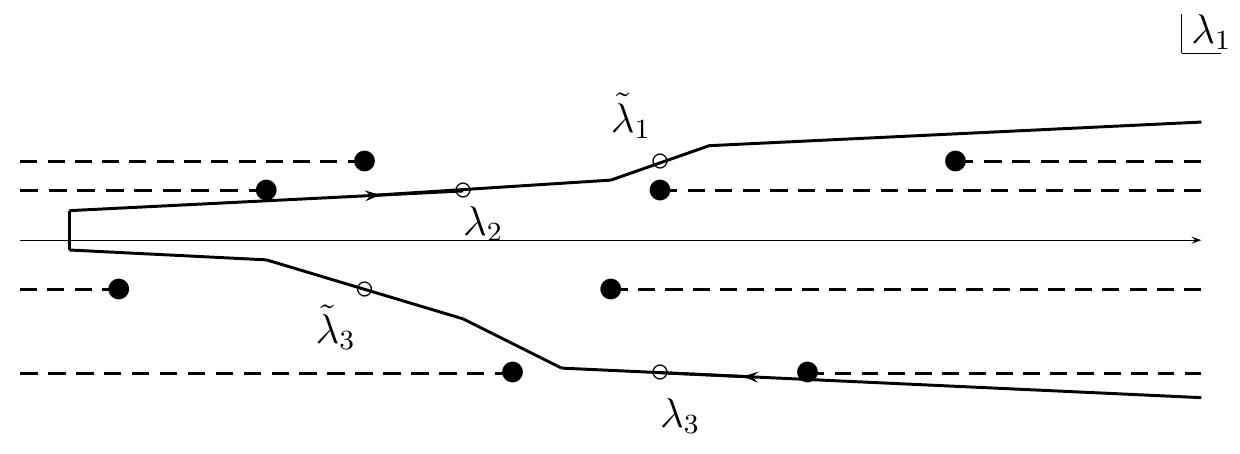}
  \end{center}
\caption{Singularities in the complex $\lambda_1$ plane.
Two external points lie at $\tilde \lambda_1, \tilde \lambda_3$ and two internal points lie at $\lambda_2,\lambda_3$ as indicated by the open circles.  Filled circles are singularities and dashed lines are branch cuts.}
\label{fig:ten}
\end{figure}

The structure of singularities in the complex $\lambda$-plane is directly analogous to that discussed in the complex $t$-plane in section \ref{factorization}.  We again fix the spatial coordinates of all points, both internal and external, and take them to be distinct.  An example for the conformal-time $\lambda_1$ of the vertex $Y_1 =(\lambda_1,\mathbf{y}_1)$ is shown in figure~\ref{fig:ten}, where dashed lines again indicate branch cuts.
Of the two singularities with the same imaginary part, we call the one with the larger (smaller) real part a past (future) singularity. For example, the singularities due to vertex $(\lambda_3,\mathbf{y}_3)$ are at
\begin{equation}
\lambda_\pm = \lambda_3 \pm \|\mathbf{y}_1-\mathbf{y}_3\|.
\end{equation}
The points $\lambda_+$ and $\lambda_-$ are a past singularity and a future singularity, respectively.  Notice that
$({\rm Re}\,\lambda_+,\mathbf{y}_1)$  and $({\rm Re}\,\lambda_-,\mathbf{y}_1)$ are on the past and future light-cones of
$({\rm Re}\,\lambda_3,\mathbf{y}_3)$, respectively.

\begin{figure}[ht]
\begin{center}
  \includegraphics{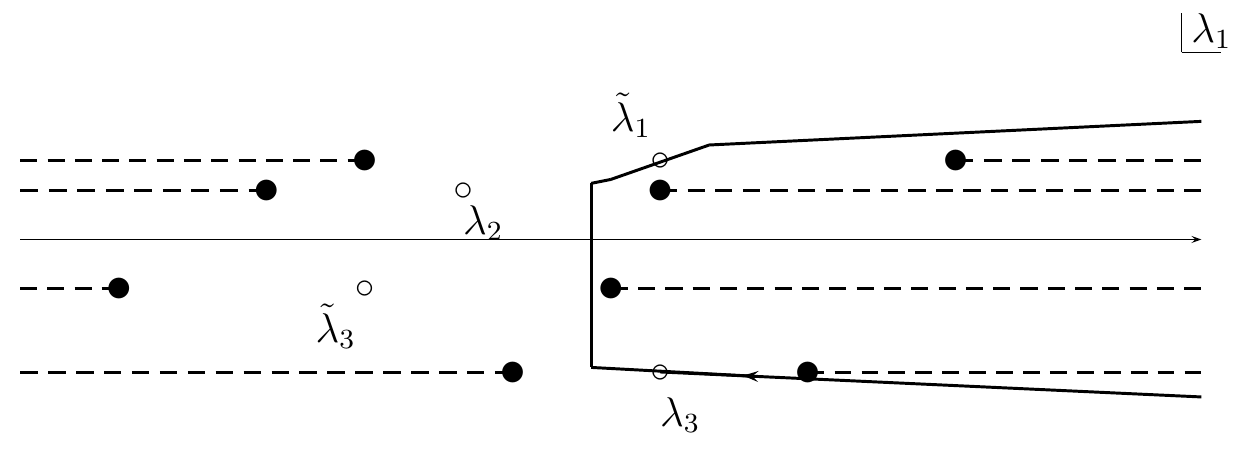}
\end{center}
\caption{The $\lambda_1$-contour in figure~\ref{fig:ten} deformed as much to the right as possible for the given values of the spatial coordinates as described in the text.}
\label{fig:eleven}
\end{figure}

Also in the same way as in section \ref{factorization}, each $\lambda_k$-contour can be deformed as we like so long as it encloses all past singularities and avoids all future singularities.   In particular, for the given values of all spatial coordinates ${\mathbf x}_i, {\mathbf y}_k$, the portion of the contour to the left of a vertical line segment connecting two points on the contour can be replaced by this line segment provided that all past singularities lie to its right.
For example, the $\lambda_1$ contour in figure~\ref{fig:ten} can be deformed as in figure~\ref{fig:eleven}.
Note that this contour may no longer pass through certain $\lambda$-values corresponding either to  external points or to other contours which were not similarly deformed.

Using this observation, we deform the contours as follows.
We begin with an integral where all $\lambda_k$ are integrated over the same contour $C$ of the form shown in figure~\ref{fig:nine} for some given values of the spatial coordinates ${\mathbf x}_j, {\mathbf y}_k$.
We deform {\em all} of the $\lambda_k$-contours {\em in the same way} as follows.
We choose a vertical line segment connecting two points of the contour such that all past singularities on the complex $\lambda_k$-plane for all $k$ is to its right. Then we let this line segment replace the portion of the contour to its left.  We keep deforming the contour in this manner by moving the vertical line segment to the right until it encounters a past singularity, say, on the $\lambda_{k_1}$-plane, $\lambda_{k_1}$ being the conformal-time for $Y_{k_1}$, due to some external point, say $X_{j_1}$.  We then stop deforming the contour for $Y_{k_1}$ (since we cannot deform it beyond the singularity) and hold it fixed.  We describe this relationship between $Y_{k_1}$ and $X_{j_1}$ by saying that $Y_{k_1}$ is past-related to $X_{j_1}$ (for the given values of all spatial coordinates) and writing\footnote{\label{tree}For certain spatial coordinates, our contour will encounter two singularities due to distinct external points $X_{j_1}$ and $X_{j_2}$ at the same time. Since this happens only on a set of spatial coordinates of measure zero, we will ignore such cases and assume below that $Y_{k_1}$ is past-related to only one point, and similarly for other vertices in the diagram.} $Y_{k_1} \to X_{j_1}$.

We then choose some value of $\lambda_{k_1}$ on its fixed contour and deform the remaining contours by moving the vertical line segment to the right with $\lambda_{k_1}$ fixed until one of them, say a contour for $Y_{k_2}$, hits a past singularity due to, say $X_{j_2}$, which is either an external point or the vertex $Y_{k_1}$ whose contour is being held fixed.  We write
$Y_{k_2}\to X_{j_2}$ and hold the contour for $Y_{k_2}$ fixed from now on. We continue in this manner until each vertex is past-related to another point, so that all contours have been fixed\footnote{\label{ldep}It may be that $X_{j_2} = Y_{k_1}$ for some values of $\lambda_{k_1}$ while for other values $X_{j_2}$ is an external point.  In this way, our definition of new past-relations can depend on the positions of integration variables along contours that have already been fixed.  It is straightforward to deal with this seeming complication as discussed in footnote \ref{forest} below.}.

\begin{figure}[ht]
\begin{center}
  \includegraphics{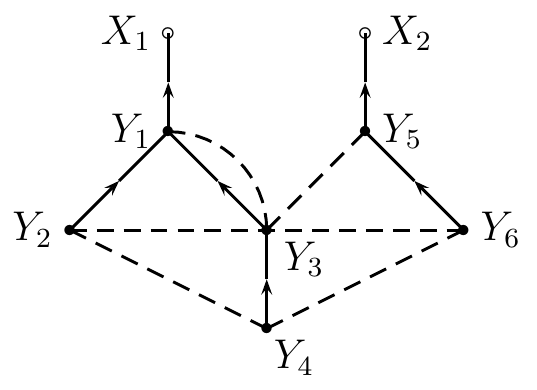}
\end{center}
\caption{A Feynman diagram in which arrows indicate past-relations as described in the text, determined by some particular set of spatial coordinates.}
\label{fig:twelve}
\end{figure}

To understand the resulting structure, we now use the above past-relations to decorate the Feynman diagram under discussion for each fixed set of spatial coordinates. Note that any pair $(A,B)$ of vertices with $A$ past-related to $B$ must be connected by at least one line on the diagram\footnote{Otherwise the location of point $B$ could not produce singularities in the propagators evaluated at $A$. In particular, our notion of past-relation is {\it not} transitive.}. If there is one line from $A$ to $B$, we decorate it with an arrow pointing from $A$ to $B$ (i.e., toward the future).  If there is more than one such line, we decorate only one of them.  Once all past-relations have been indicated in this way, we replace all remaining undecorated propagators with dashed lines.  An example is shown in figure~\ref{fig:twelve}.

In the deformation of contours described above, the contours are deformed until one of them encounters a past singularity.  Although this procedure is sufficient to show the convergence of ${\mathcal A}_P$ itself, we need to modify it slightly for proving convergence of the derivatives of ${\mathcal A}_P$ with respect to the external coordinates, which diverge at past singularities. Here we briefly describe this modification. The main difference is that the modified deformation keeps the contours away from past singularities.

We choose the initial contour common to all $\lambda_k$ as before.
We define an \emph{effective past singularity} as follows: if $\lambda$ is a past singularity, then the corresponding effective past singularity is $\lambda - b$, where $b$ is a small but positive constant.  We deform the contours in the same way as before except that they are deformed until one of the contours encounters an effective past singularity rather than a true one.  We define the past-relation as before.
It may happen that some effective past singularities are outside the contour though the true ones must be inside.  If the effective past singularity on the complex $\lambda_{k_3}$-plane due to a point $X_{j_3}$, external or internal, is outside the contour, we stop deforming the contour for $\lambda_{k_3}$, fix the value of $\lambda_{k_3}$ on this contour, and let $Y_{k_3} \to X_{j_3}$. (If there are two or more effective past singularities outside the contour, we choose one to define the past-relation.)  The rest is the same as the original deformation\footnote{Notice that, since the contour is separated from the past singularities due to external points by
a finite distance
for large $|\lambda|$, the $Z$ in (\ref{invP}) for an external propagator is bounded away from $1$ as $|\lambda|\to \infty$.  This means that a differentiated external propagator is bounded on the contours and that the proof for convergence of $\mathcal{A}_P$ below can be used virtually unaltered for the derivative of $\mathcal{A}_P$.}.

As noted in the introduction, it is well-known that in-in diagrams can be computed by integrating only over the past light cones of external points.  The choice of contours above gives a similar result, but one which is clearly valid for finite $\epsilon$.  To see the similarity, note that when $A$ is past-related to $B$ the real part of point $A$ lies in the casual past of the real part of point $B$ over most of the contour for $A$.  The exception is a finite piece near the minimum value of ${\rm Re}\ \lambda$  due to the use of effective past singularities in the modified contour
deformation.  Since any internal point is connected by some chain of arrows to some external point, except for a set of finite-sized pieces as noted above, the projection of the integration region onto the real $\lambda$-axis lies in the causal past of at least one external point. We will find it useful below to break up the integration region into such past light cones and finite-sized protruding segments.

Now we establish convergence using the modified deformation of contour.  Recall that each internal point $A$ is past-related to precisely one point $B$ (see footnote \ref{tree}), which may be either internal or external.  Also recall that, starting at any internal point, one may always follow a chain of arrows upwards until one arrives at an external point.  As a result, deleting all dashed lines results in a set of disconnected subdiagrams for which each connected component is a tree whose root (which in this case means that future-most point) is an external point.  As a result, if we replace every dashed-line propagator by the bound $B(M)$, our amplitude ${\mathcal A}_P$ factorizes into a product of tree amplitudes in which all points are connected by a chain of past-relations\footnote{\label{forest} Since past-relations depend on both the spatial coordinates and the conformal time coordinates of the previously-fixed contours (see footnote \ref{ldep}),   the tree structure exhibits a similar dependence. It would therefore be better to say that each amplitude can be written as a finite sum of products of tree amplitudes, where the amplitudes for any given term in the product are integrated only over some subset of the spacetime coordinates.  But since we wish only to establish absolute convergence of the amplitude, it does no harm to extend the spatial integrations for each tree to the full space ${\mathbb R}^{n(D-1)}$ and each $\lambda$-integrations over the whole of the appropriate contour and to then abuse language by referring to the amplitude as a `product' of tree amplitudes without mentioning the remaining sum explicitly.}.

But each such tree amplitude is easy to bound.  We begin by bounding the integrals corresponding to some past-most vertex $Y = (\lambda, {\mathbf y})$ in a given tree (e.g., $Y_2,$ $Y_4$ or $Y_6$ in figure~\ref{fig:twelve}).  Taking the magnitude of the integrand, this integral takes the form
\begin{equation}
I = \int d^{D-1}\mathbf{y}\int_{C_{\mathbf y}} \frac{|d\lambda|}{|\lambda|^D}|\Delta^{\rm reg}(Y',Y)|, \label{Istart}
\end{equation}
where the notation indicates that the contour $C_{\mathbf y}$ over which we integrate $\lambda$ can depend on the spatial coordinates ${\mathbf y}$.  Recall that $\Delta^{\rm reg}(Y',Y)$ behaves like $|Z|^{-\nu}$, $\nu>0$, for large $Z = Y \cdot Y'$.  As a result, $|\Delta^{\rm reg}(Y',Y)|$ behaves at most like $Z_R^{-\nu}$, where $Z_R:= {\rm Re}\ Z$.  Let us choose some $Z_0$ large enough that for $Z_R > Z_0$ our $|\Delta^{\rm reg}(Y',Y)|$ is bounded by $\alpha Z_R^{-\nu}$ for some real constant $\alpha$.

It is now useful to break up the integration domain into several pieces.  First consider the portion of $C_{\mathbf{y}}$ noted above that protrudes from the past light cone of ${\rm Re}\ Y'$.   The past singularity which has stopped this contour from being deformed further is
at $\lambda^{+}_y = \lambda_{y'} + \|\mathbf{y}-\mathbf{y'}\|$ and the corresponding effective past singularity is at
$\lambda^{+,{\rm eff}}_y = \lambda^{+}_y - b$  on the complex $\lambda_{y}$-plane. Hence the length of this portion of the contour is bounded by
a constant, which is larger than $2b$ because the contour has a finite width.
We also find that ${\rm Re}\ \lambda_y \geq \lambda_0 + c\|\mathbf{y}-\mathbf{y}'\|$, where $\lambda_0$ and $c$ are some positive constants, on this portion of the contour.  This is because
${\rm Re}\ \lambda_y \geq \lambda_{\rm min}$, where $\lambda_{\rm min}$ is the minimum of the real part of $\lambda_y$ at $\mathbf{y}=\mathbf{y}'$, and that ${\rm Re}\,\lambda^{+,{\rm eff}}_y/\|\mathbf{y}-\mathbf{y'}\|\to 1$ as $\|\mathbf{y}\|\to \infty$.
The contribution to $I$ from the protruding portions is thus bounded by a constant times
$B(M)\int d^{D-1}\mathbf{y}(\lambda_{0}+c\|\mathbf{y}\|)^{-D}$.

Next consider the contribution to $I$ from the region $0 < Z_R < Z_0$.  This is bounded by $\beta B(M)$ times the total measure $\int d^{D-1}\mathbf{y}\int |d\lambda|\,|\lambda|^{-D}$  of this region, where $\beta$ is a constant,  assuming that this measure is finite.  To see that this is so, consider any point $X$ in the Poincar\'e patch of real de Sitter space and, furthermore, consider  the part of its past light cone that is both within embedding distance $Z_0$ and which also lies to the future of the cosmological horizon. This region  is compact and thus has finite volume.  Since widening of the contour described above for large ${\rm Re}\ \lambda$ has little effect at large $Z$, for fixed $\epsilon$, we may therefore choose $Z_0$ large enough that the measure of the desired region in complex de Sitter is within, say, a factor of $2$ of the volume of the region just discussed in real de Sitter space.  Thus this part of our integral is easily bounded.

We can similarly bound the contribution from the region $Z_R > Z_0$.  For large enough $Z_0$, this contribution is no more than, say, a factor of 2 times the integral of $\alpha Z_R^{-\nu}$ over the region of real de Sitter space lying to the future of the cosmological horizon but more than an embedding distance $Z_0$ to the past of the point $({\rm Re}\ \lambda_{y'},\mathbf{y}')$. To proceed further, one should compute the volume of surfaces lying a constant embedding distance $Z_R$ to the past of the given point but to the future of the cosmological horizon.  In the limit of large $Z_R$, this volume turns out to approach the constant $1/(D-1)$.  We also note that the proper time difference between the two surfaces at $Z_R$ and $Z_R+dZ_R$ is $dZ_R/Z_R$ for large $Z_R$. As a result, for large enough $Z_0$ the contribution from the region ${\rm Re \ Z} > Z_0$ is bounded by, say,
$4 \alpha (D-1)^{-1} \int_{Z_0}^{\infty} dZ_R Z_R^{-1-\nu}$. Combining this with our observations above shows that (\ref{Istart}) is
bounded by some constant $B(I)$ which (for, say, $|\epsilon| < 1$) depends only on  the mass of our field and which in particular is independent of both $\epsilon$ and the location of the point $Y'$.

As a result, we can bound the integral corresponding to any of the above tree diagrams by $B(I)$ times the integral corresponding to the diagram shortened  by cutting off a lowest line.  We can clearly repeat this procedure and continue to remove the lowest lines until we are left with no lines at all.  Thus, the integral corresponding to each arrowed tree diagram is bounded by $(B(I))^n$, where $n$ is the number of lines in the given tree.  Hence the integral for the amplitude $\mathcal{A}_P$ given by (\ref{ininA}) is (absolutely) convergent after translating the contours appropriately.

\subsection{Analyticity of the amplitude ${\mathcal A}_{P}$}
\label{regA}

To complete the argument for equivalence between the Euclidean and Poincar\'e in-in correlators, we now establish the desired analyticity property of the amplitude ${\mathcal A}_P$, which we have shown above to be well-defined. Specifically, we will show that $\mathcal{A}_P$ is analytic as a function of the conformal times $\tilde\lambda_i$ of the external points if $(\tilde\lambda_1,\ldots,\tilde\lambda_m)\in U = \{(\mu_1,\ldots,\mu_m)\in \mathbb{C}^m: {\rm Im}\ \mu_i < {\rm Im}\ \mu_{i+1}, i=1,\ldots,m-1\}$, or more generally if the imaginary parts of $\tilde\lambda_i$ are all distinct, for any given spatial coordinates $\mathbf{x}_i$.
For this purpose we introduce an additional regulator defined by some $s > 0$ and show that the regulated correlators are analytic functions on $U$.  We then show that this analyticity property persists in the $s\to 0$ limit.

Our choice of regulator is straightforward to introduce.
We define the amplitude $\mathcal{A}_{P,s}$ for $s > 0$ by simply replacing each (already Pauli-Villars regulated) propagator $\Delta^{\rm reg}(Z)$ with $\Delta^{\rm reg}_s(Z) = \Delta^{\rm reg}(Z-s)$, where these propagators are written as functions of the embedding distance $Z$ defined by (\ref{invP}). Note that $s$ is indeed a regulator in the sense that it widens the gap between any pair of past and future singularities such as those shown in figures~\ref{fig:ten} and \ref{fig:eleven}.  As a result, any contour that can also be used to compute the unregulated ${\mathcal A}_{P}$ can be used to compute ${\mathcal A}_{P,s}$ for $s > 0$. Thus, contours similar to figure~\ref{fig:nine} are again allowed for $(\tilde\lambda_1,\ldots,\tilde\lambda_m)\in U$.  The (absolute) convergence of the integrals for ${\mathcal A}_{P,s}$ can be established in exactly the same way as in the $s=0$ case.

Now consider complex $\tilde{\lambda}_i$-derivatives of ${\mathcal A}_{P,s}$ computed formally by differentiating the integrand, which is a product of propagators, and then integrating over the contours.  Our $s$-regularization makes the integrand analytic in an open neighborhood of $(\tilde\lambda_1,\ldots,\tilde\lambda_m)$ with the contours fixed so that complex derivatives of the integrand are well-defined.  Furthermore, differentiated propagators are bounded at fixed $s$ and their behavior as $Z\to \infty$ is not worse than that of un-differentiated propagators.  Hence the argument for the (absolute) convergence of the integrals defining ${\mathcal A}_{P,s}$ applies equally well to integrals of the differentiated integrands.  But absolute convergence guarantees that these latter integrals do in fact give the complex $\tilde{\lambda}_i$-derivatives of ${\mathcal A}_{P,s}$.  It follows that such integrals are well-defined and that each ${\mathcal A}_{P,s}$ is analytic in $U$.

Now, since the integrals defining ${\mathcal A}_P$ converge, it is clear that ${\mathcal A}_{P,s}$ tends to ${\mathcal A}_P$ as $s\to 0$.
As for the $\lambda_i$-derivative of ${\mathcal A}_{P,s}$, the integrand will be divergent in the $s\to 0$ limit only where the arguments of the differentiated external (regulated) propagator, become coincident.  However, due to our Pauli-Villars regularization this divergence is very mild and does not spoil absolute convergence.  It follows that the $\lambda_i$-derivative of ${\mathcal A}_{P,s}$ has a finite limit as $s\to 0$ which gives the $\lambda_i$-derivative of ${\mathcal A}_{P}$.  In particular, these derivatives are well-defined on $U$, so that ${\mathcal A}_{P}$ is analytic in this domain.  This completes our step 3.

Let us now assemble the facts demonstrated above to establish the equivalence of the Euclidean and Poincar\'e in-in correlators.
The amplitude ${\mathcal A}_P$ and the corresponding Euclidean amplitude, which we call ${\mathcal A}_E$, are both analytic functions of the conformal-time variables
$\tilde{\lambda}_i$ of the external points $(\tilde{\lambda}_i,\mathbf{x}_i)$ if $(\tilde\lambda_1,\ldots,\tilde\lambda_m)\in U$ (Step 3).  These amplitudes coincide in the limit where the imaginary parts of the conformal-time variables tend to zero if the limits of the external points all lie in the static patch of real de~Sitter space. (This was established in two steps: In step 1 we established that ${\mathcal A}_E$ agrees with the static in-in amplitude, and in step 2 we established (rather trivially) that the latter agrees with the Poincar\'e in-in amplitude if the limits of the external points are all in the static patch of real de~Sitter space.)  Hence, by uniqueness of analytic continuation\footnote{Here, we are using the agreement of ${\mathcal A}_P$ and ${\mathcal A}_E$ on an open subset of a real section, $B=\{(\mu_1,\ldots,\mu_m)\in \mathbb{C}^m:{\rm Im}\ \mu_i=0, i=1,\ldots,m\}$, on the boundary of the region of analyticity $U$ to conclude ${\mathcal A}_P={\mathcal A}_E$ in $U$.  This is a simple corollary of Bogolubov's edge-of-the-wedge theorem (see, e.g., Theorem 2-17 in~\cite{Streater:1989vi}).}, ${\mathcal A}_P = {\mathcal A}_E$ for all $\tilde\lambda_i$ wherever these amplitudes are well-defined.  Then, ${\mathcal A}_P$ and ${\mathcal A}_E$ have, of course, the same limit as ${\rm Im}\ \tilde\lambda_i\to 0$, producing the same physical amplitude for any points $X_i$ in the Poincar\'e patch.

\section{Explicit checks in simple examples}
\label{numerics}

As a check on our arguments, we now explicitly compare the Euclidean and Poincar\'e in-in results for one-loop corrections to propagators from $\phi^4$ and $\phi^3$ interactions.  As the Euclidean computations (including the analytic continuation
to Lorentz-signature de Sitter) were performed in \cite{Marolf:2010zp}, we focus on the in-in calculations here.

%%%%%%%%%%%%%%%%%%%%%%%%%%%%%%%%%%%%%%%%%%%%%%%%%%%%%%%%%%%%%%%%%
\subsection{$\phi^4$ correction}
\label{sec:phi4}
%%%%%%%%%%%%%%%%%%%%%%%%%%%%%%%%%%%%%%%%%%%%%%%%%%%%%%%%%%%%%%%%%

\begin{figure}[t]
  \centering
   \includegraphics{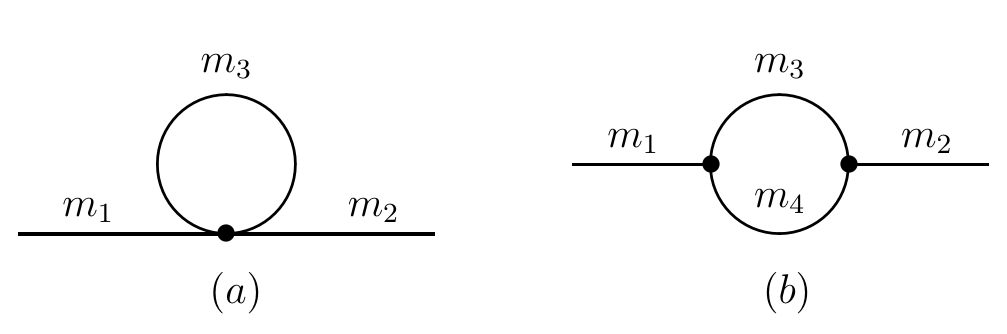}
   \caption{The 1-loop corrections to the propagator.}
   \label{fig:1loop}
\end{figure}

Consider the 1-loop correction to the propagator
due to an interaction term of the type
$\mathcal{L}_{\rm int}[\phi] = -\frac{\lambda}{4!} \phi(X)^4$.
The relevant Feynman diagram is shown in Fig.~\ref{fig:1loop} (a).
The in-in correlation function is given by
\eqn{ \label{eq:phi4InIn}
  \C{T \phi(X_1)\phi(X_2)}
  &=& - \frac{\lambda i}{2} \int_Y
  \bigg\{ \D^{\rm reg}_{m^2_3}(Y,Y) \D_{m^2_1}(Y,X_1) \D_{m^2_2}(Y,X_2)
  \nn \\ & & \phantom{\frac{\lambda i}{2} \int_Y \bigg\{ \;}
  - \D^{{\rm reg}\,*}_{m^2_3}(Y,Y) W_{m^2_1}(Y,Y_1) W_{m^2_2}(Y,X_2) \bigg\}
}
Here $\int_Y\dots$ denotes an integral over the Poincar\'e patch,
and we remind the reader that $\D_{m^2}(X,Y)$, $\D_{m^2}^*(X,Y)$, and
$W_{m^2}(X,Y)$ are the time-ordered, anti-time-ordered, and Wightman
2-point functions of the Gaussian theory. It is convenient to let
each line in the Feynman diagram have a distinct mass; one may take
the limit of equal masses later.
This expression has a UV divergence for $D \ge 4$ which we control by using Pauli-Villars regularization.  For simplicity, we regulate only the internal lines, though we could of course also regulate the external lines as well.

To simplify (\ref{eq:phi4InIn}) we first note that the regulated Feynman
function $\D_{m^2}^{\rm reg}(X,Y)$ evaluated at coincident points is real
and independent of position, so
$\D_{m^2}^{\rm reg}(Y,Y) = \D_{m^2}^{{\rm reg}\,*}(Y,Y) =: \D_{m^2}^{\rm reg}(1)$.
After removing a common factor of
$\D_{m^2_3}^{\rm reg}(1)$ from the integrand the remaining integral is
\eq{ \label{eq:I}
  I(X_1,X_2) := i \int_Y
  \left\{ \D_{m^2_1}(Y,X_1) \D_{m^2_2}(Y,X_2)
    - W_{m^2_1}(Y,X_1) W_{m^2_2}(Y,X_2) \right\} .
}

The integral (\ref{eq:I}) can be quickly performed as follows.
Consider a theory of two free massive scalar fields
$\Phi_{1,2}(X)$ with masses $M_1^2 \neq M_2^2$.
We can re-write this theory in terms of two new fields $\phi_{1,2}(X)$ by
performing an $SO(2)$ rotation in field space:
\eqn{ \label{eq:rotation}
  \phi_1(X) &=& \cos\omega \,\Phi_1(X) - \sin\omega \,\Phi_2(X), \nn \\
  \phi_2(X) &=& \sin\omega \,\Phi_1(X) + \cos\omega \,\Phi_2(X) .
}
The fields $\phi_{1,2}(X)$ have masses $m_{1,2}^2$ that are
functions of $M_{1,2}^2$ and $\omega$, and also an interaction
$-g \phi_1(X)\phi_2(X)$ in the Lagrangian with the coupling
$g=(M_1^2-M_2^2)\sin \omega\cos\omega$.
Now consider the correlation function $\C{T \phi_1(X_1)\phi_2(X_2)}$.
We may compute this correlation function using standard in-in
perturbation theory; the term at lowest order in $g$ (or equivalently, in $\omega$) is
\eq{ \label{eq:gI}
  \C{T \phi_1(X_1)\phi_2(X_2)} = - g I(X_1,X_2) + O(g^3).
}
On the other hand, by
simply using (\ref{eq:rotation})
we can compute $\C{T \phi_1(X_1)\phi_2(X_2)}$
exactly\footnote{A truly skeptical reader might ask whether (\ref{eq:gI}) must necessarily give the vacuum correlator of the theory defined by (\ref{eq:rotation}).  But at this order the result must be a Gaussian state invariant under translations, rotations, and the scaling symmetry of the Poincar\'e patch.  This determines the state uniquely, assuming that the results are finite.   Finiteness in turn can be shown by either a careful direct analysis or by using the results of \cite{Higuchi:2009ew} to expand both $m_1^2$ and $m_2^2$ about the conformal coupling value $m_c^2 = \frac{1}{4}D(D-2)$ and then using the explicit calculations of that reference.} :
\eqn{
  \label{eq:trick}
  \C{T \phi_1(X_1)\phi_2(X_2)} &=& \sin\omega \cos\omega
  \big[ \C{T\Phi_1(X_1)\Phi_1(X_2)} - \C{T\Phi_2(X_1)\Phi_2(X_2)} \big]
  \nn \\
  &=& \sin\omega \cos\omega \big[
  \D_{M_1^2}(X_1,X_2) - \D_{M_2^2}(X_1,X_2) \big] .
}
We can then write $M_1^2$, $M_2^2$ and $\omega$ in terms of
$m_1^2$, $m_2^2$ and $g$, expand the right-hand side of
(\ref{eq:trick}) in a power series in $g$, and equate
the $O(g)$ term with the right-hand side of (\ref{eq:gI}).
The result is the equality
\eq{ \label{eq:Ians}
  I(X_1,X_2)
  = \frac{\D_{m^2_1}(X_1,X_2) - \D_{m^2_2}(X_1,X_2)}{m^2_2 - m^2_1} .
}

Returning to (\ref{eq:phi4InIn}), we may use (\ref{eq:Ians}) to obtain
\eq{ \label{eq:phi4LFinal}
  \C{T\phi(X_1)\phi(X_2)}
  = \frac{\lambda}{2} \D^{\rm reg}_{m^2_3}(1)
  \left[
    \frac{\D_{m^2_1}(X_1,X_2) - \D_{m^2_2}(X_1,X_2)}{m^2_1 - m^2_2}
  \right] .
}
It is clear that the same steps can be used to compute the Euclidean expression.  The analogue of (\ref{eq:I}) then involves only Euclidean propagators, but these are just what are needed to arrive at the analogue of (\ref{eq:Ians}).  After analytic continuation to real de Sitter space, the result is precisely (\ref{eq:phi4LFinal}).

We note that the above calculations
could be performed equally well using dimensional regularization
rather than the Pauli-Villars scheme. In dimensional regularization
the computation is performed in an arbitrary real dimension
which is sufficiently small such that there are no ultraviolet
divergences. As in Pauli-Villars regularization, the values
of de Sitter-invariant Green's functions $\D_{m^2}(1)$, etc., are
divergent but de Sitter-invariant constants.
By the usual arguments \cite{Collins:1984xc}, the
manipulations we performed to derive
(\ref{eq:Ians}) and its Euclidean analogue are valid for arbitrary real
dimension.

%%%%%%%%%%%%%%%%%%%%%%%%%%%%%%%%%%%%%%%%%%%%%%%%%%%%%%%%%%%%%%%%%
\subsection{$\phi^3$ correction}
\label{sec:phi3}
%%%%%%%%%%%%%%%%%%%%%%%%%%%%%%%%%%%%%%%%%%%%%%%%%%%%%%%%%%%%%%%%%

Next we turn to the 1-loop correction to the propagator
that arises from the interaction
$\mathcal{L}_{\rm int}[\phi] = - \frac{g}{3!} \phi^3(X)$. The
relevant Feynman diagram is shown in Fig \ref{fig:1loop} (b). Once
again it is convenient to let each leg of this diagram have a
distinct mass. This correction has a UV divergence for spacetime
dimension $D \ge 4$. Both to draw on results of \cite{Marolf:2010zp} and to simplify the arguments, we carry out the computations below using dimensional regularization.
In particular, we will compute this correction in
arbitrary $D < 2$, then analytically continue $D$ to extend the result
to higher dimensions. However, we also explain how similar results (with more complicated explicit forms) can be obtained via Pauli-Villars techniques.

It is useful to introduce a so-called linearization formula for the Green's
functions $\D_{m^2}(X,Y)$, $\D_{m^2}^*(X,Y)$ and $W_{m^2}(X,Y)$. We use the variable $\a := (D-1)/2$ to keep track of spacetime dimension and the
mass variable $\s$ defined by the equation
$-\s(\s+2\a) = m^2\ell^2$. All three
Green's functions are proportional to the Gegenbauer function
$C^{\a}_\s(Z)$. The following linearization formula for the
Gegenbauer function allows us to replace a product of Gegenbauer
functions with an integral of a single Gegenbauer function
\cite{Marolf:2010zp}:
\eq{ \label{eq:linearization}
  C_{\s_1}^\a(Z)C_{\s_2}^\a(Z)
  = - \frac{4\pi^\a}{\Gamma(\a)} \sin(\pi\s_1) \sin(\pi\s_2)
  \int_{\mu} \frac{\rho^\a_{\s_1\s_2}(\mu)}{\sin(\pi\mu)}
  C_\mu^\a(Z) .
}
In this equation $C^\a_\s(Z)$ is the Gegenbauer function which
is analytic in the complex $Z$ plane cut along $Z \in (-\infty,-1]$.
We assume ${\rm Re\,}\s_1 < 0$ and ${\rm Re\,}\s_2 < 0$, which is
valid for $m_{1,2}^2 > 0$.
The shorthand $\int_\mu \dots$ denotes a contour integral in the complex
$\mu$ plane with measure $d\mu/2\pi i$. The integration contour runs
from $-i\infty$ to $+i\infty$ within the strip
${\rm Re}(\s_1+\s_2) < {\rm Re}\,\mu < 0$. Within this strip the integrand
is analytic and the contour integral converges absolutely. From
(\ref{eq:linearization}) we may write the following linearization
formula for the Green's functions, with $H_\s(X,Y)$ standing
for $\D_\s(X,Y)$, $\D_\s^*(X,Y)$, or $W_\s(X,Y)$:
\eq{ \label{eq:Hlin}
  H_{\s_1}(X,Y)H_{\s_2}(X,Y) = \int_\mu \rho^\a_{\s_1\s_2}(\mu) H_\mu(X,Y) .
}

Of course, most of the content of (\ref{eq:Hlin}) is contained
in the details of the function $\rho^\a_{\s_1\s_2}(\mu)$. The explicit
form of $\rho^\a_{\s_1\s_2(\mu)}$ can
be found in \cite{Marolf:2010zp}\footnote{The definition of
$\rho^\a_{\s_1\s_2}(\mu)$ used here is $(-2)$ times the
$\rho^\a_{\s_1\s_2}(L)$ of that paper.};
we will not need the explicit form. We need only note that:
\begin{enumerate}
\item
  $\rho^\a_{\s_1\s_2}(\mu)$ is itself analytic in the region
  ${\rm Re\,} \mu > {\rm Re}(\s_1+\s_2)$ and that in this region the
  function behaves at large $|\mu|\gg 1 $ like $|\mu|^{2\a-3}\log(\mu)$.
  In particular, it follows that
  \eq{ \label{eq:muInt}
    \int_\mu \frac{\rho^\a_{\s_1\s_2}(\mu)}{m^2 + \mu(\mu+2\a)} = 0
    \quad {\rm for}\; \a < 2
  }
  for $m^2 > 0$ with the $\mu$ contour lying to the right of poles
  at $\mu = -\a \pm \sqrt{\a^2 - m^2}$ (both of which lie in the
  left half-plane).
\item
  The function $\rho^\a_{\s_1\s_2}(\mu)$ is proportional to $\Gamma(2-2\a)$
  and so
  has simple poles as a function of $\a$ at $\a = 1,3/2,2,\dots$.
  Of course, the left-hand sides of (\ref{eq:linearization}) and
  (\ref{eq:Hlin}) are regular for these values of $\a$; the integral
  over $\mu$ cancels these poles. However, the integral of an
  arbitrary function of $\mu$ times $\rho^\a_{\s_1\s_2}(\mu)$ will
  generically not cancel this divergence and so will diverge
  at these values of $\a$.
\end{enumerate}

The $O(g^2)$ correction to the propagator in this theory is given
in the in-in formalism by the expression
\eqn{ \label{eq:phi3InIn}
  \left\langle T \phi(X_1)\phi(X_2) \right\rangle
  &=& - g^2 \int_{Y_1}\int_{Y_2} \bigg\{
    \D_{\s_1}(Y_1,X_1) \D_{\s_2}(Y_2,X_2) \D_{\s_3}(Y_1,Y_2) \D_{\s_4}(Y_1,Y_2)
    \nn \\ & & \phantom{g^2 \int_{Y_1}\int_{Y_2} \bigg\{ \;}
  - W_{\s_1}(Y_1,X_1) \D_{\s_2}(Y_2,X_2) W_{\s_3}(Y_1,Y_2) W_{\s_4}(Y_1,Y_2)
    \nn \\ & & \phantom{g^2 \int_{Y_1}\int_{Y_2} \bigg\{ \;}
  + W_{\s_1}(Y_1,X_1) W_{\s_2}(Y_2,X_2) \D^*_{\s_3}(Y_1,Y_2) \D^*_{\s_4}(Y_1,Y_2)
    \nn \\ & & \phantom{g^2 \int_{Y_1}\int_{Y_2} \bigg\{ \;}
  - \D_{\s_1}(Y_1,X_1) W_{\s_2}(Y_2,X_2) W_{\s_3}(Y_2,Y_1) W_{\s_4}(Y_2,Y_1)
  \bigg\}  . \nn \\
}
The first two terms in (\ref{eq:phi3InIn}) contain the integral
over $Y_1$:
\eqn{
  T_1 &:=& \int_{Y_1} \bigg\{
    \D_{\s_1}(Y_1,X_1) \D_{\s_3}(Y_1,Y_2) \D_{\s_4}(Y_1,Y_2)
    \nn \\ & & \phantom{\int_{Y_1} \bigg\{ \;}
  - W_{\s_1}(Y_1,X_1) W_{\s_3}(Y_1,Y_2) W_{\s_4}(Y_1,Y_2)
  \bigg\} .
}
To compute $T_1$ we first use the linearization formula
(\ref{eq:Hlin}) in each term,  then use (\ref{eq:Ians}) to
integrate over $Y_1$:
\eqn{ \label{eq:T1}
  T_1 &=& \int_\mu \rho^\a_{\s_3\s_4}(\mu)
  \int_{Y_1} \left\{
    \D_{\s_1}(Y_1,X_1)\D_{\mu}(Y_1,Y_2)
    - W_{\s_1}(Y_1,X_1)W_{\mu}(Y_1,Y_2) \right\}
  \nn \\
  &=& \frac{1}{i} \int_\mu \rho^\a_{\s_3\s_4}(\mu)
  \left[ \frac{\D_{\mu}(Y_2,X_1) - \D_{\s_1}(Y_2,X_1)}{m_1^2-m_\mu^2}  \right]
  \nn \\
  &=& \frac{1}{i} \int_\mu \frac{\rho^\a_{\s_3\s_4}(\mu)}{m_1^2-m_\mu^2}
  \D_{\mu}(Y_2,X_1) .
}
We compute with $\a < 3/2$,
so the final equality follows from (\ref{eq:muInt}). The latter two
terms in (\ref{eq:phi3InIn}) contain the integral over $Y_1$:
\eqn{ \label{eq:T2tmp}
  T_2 &:=& \int_{Y_1} \bigg\{
  W_{\s_1}(Y_1,X_1) \D^*_{\s_3}(Y_1,Y_2) \D^*_{\s_4}(Y_1,Y_2)
  \nn \\ & & \phantom{\int_{Y_1} \bigg\{ \;}
  - \D_{\s_1}(Y_1,X_1) W_{\s_3}(Y_2,Y_1) W_{\s_4}(Y_2,Y_1)
  \bigg\} .
}
To compute $T_2$ we again use the linearization formula
(\ref{eq:Hlin}), then use the integral
\eqn{ \label{eq:J}
  J(X_1,X_2) &:=& i \int_Y
  \left\{ W_{\s_1}(X_1,Y) \D_{\s_2}(Y,X_2)
    - \D^*_{\s_1}(X_1,Y) W_{\s_2}(Y,X_2) \right\}
  \nn \\
  &=&
  \frac{ W_{\s_1}(X_1,X_2) - W_{\s_2}(X_1,X_2) }{ m_2^2 - m_1^2 } .
}
This integral may be derived in the same manner as $I(X_1,X_2)$ by
examining the Wightman correlation function
$\C{\phi_1(x_1)\phi_2(x_2)}$ in the $SO(2)$-rotated theory.

Inserting (\ref{eq:J}) into (\ref{eq:T2tmp}) yields
\eqn{ \label{eq:T2}
  T_2 &=& \int_\mu \rho^\a_{\s_3\s_4}(\mu)
  \int_{Y_1} \left\{
    W_{\s_1}(Y_1,X_1) \D^*_{\mu}(Y_1,Y_2)
    - \D_{\s_1}(Y_1,X_1) W_{\mu}(Y_2,Y_1) \right\}
  \nn \\
  &=& - \frac{1}{i} \int_\mu \rho^\a_{\s_3\s_4}(\mu)
  \left[ \frac{ W_\mu(Y_2,X_1) - W_{\s_1}(Y_2,X_1) }{m_1^2 - m_\mu^2} \right]
  \nn \\
  &=& - \frac{1}{i} \int_\mu \frac{\rho^\a_{\s_3\s_4}(\mu)}{m_1^2 - m_\mu^2}
  W_{\mu}(Y_2,X_1) .
}
Once again the last equality follows from (\ref{eq:muInt}).
Assembling (\ref{eq:T1}) and (\ref{eq:T2}) we may write the
propagator correction as
\eqn{
  \left\langle T \phi(X_1)\phi(X_2) \right\rangle
  &=&
  - \frac{g^2}{i} \int_\mu \frac{\rho^\a_{\s_3\s_4}(\mu)}{m_1^2 - m_\mu^2}
  \int_{Y_2} \big\{
    \D_{\s_2}(Y_2,X_2)\D_\mu(Y_2,X_1)
    \nn \\ & & \phantom{ - \frac{g^2}{i} \int_\mu \frac{\rho^\a_{\s_3\s_4}(\mu)}{m_1^2 - m_\mu^2}
  \int_{Y_2} \big\{ ; }
    - W_{\s_2}(Y_2,X_2)W_{\mu}(Y_2,X_1)
  \big\} . \nn \\
}
The remaining integral over $Y_2$ may be performed using (\ref{eq:Ians}):
\eqn{ \label{eq:phi3InInFinal}
  \left\langle T \phi(X_1)\phi(X_2) \right\rangle
  &=& g^2 \int_\mu \frac{\rho^\a_{\s_3\s_4}(\mu)}{m_1^2 - m_\mu^2}
  \left[ \frac{\D_\mu(X_1,X_2) - \D_{\s_2}(X_1,X_2)}{m_2^2 - m_\mu^2} \right]
  \nn \\
  &=& g^2 \int_\mu \frac{\rho^\a_{\s_3\s_4}(\mu)}{(m_1^2 - m_\mu^2)(m_2^2 - m_\mu^2)}
  \D_\mu(X_1,X_2) .
}
The expected UV divergence of this expression is in the factor
$\Gamma(2-2\a)$ contained in $\rho^\a_{\s_1\s_2}(\mu)$.

The Euclidean computation is essentially identical, using the analogue of (\ref{eq:J}) involving only Euclidean propagators\footnote{The Euclidean analogue of (\ref{eq:J}) is identical to the Euclidean analogue of (\ref{eq:I}).}, so that the results agree under analytic continuation as desired.  The details of the Euclidean calculation were given in \cite{Marolf:2010zp}, where it is also shown that both the final expression and the counterterms used to render a finite expression in higher
dimensions agree with the standard flat-space results in
the limit $\ell\to\infty$.

One can perform essentially the same computations using Pauli-Villars regularization instead of dimensional regularization.  Note that the key steps above were the linearization formula (\ref{eq:linearization}), the property (\ref{eq:muInt}) of the form factor $\rho^{\alpha}_{\sigma_1\sigma_2}$, and the composition rules (\ref{eq:Ians}) and (\ref{eq:J}).   But it is clear from the derivation in \cite{Marolf:2010zp} that a similar linearization formula can be used to express the product of two Pauli-Villars regularized Green's functions as an integral over (un-regularized) Gegenbauer functions.  In this case, the corresponding form factor $\rho^{\alpha,M}_{\sigma_1\sigma_2}$ is manifestly finite for all $\alpha$, but depends on the Pauli-Villars regulator mass $M$.  While $\rho^{\alpha,M}_{\sigma_1\sigma_2}$ is analytic as above, it falls off faster at large $\mu$ so that the analogue of (\ref{eq:muInt}) is in fact satisfied for all $\alpha$.  Expanding any remaining regularized propagators as a sum of un-regularized propagators then allows us to apply the composition rules (\ref{eq:Ians}) and (\ref{eq:J}) and to complete the calculation.  The result is similar to that above with the replacement $\rho^{\alpha}_{\sigma_1\sigma_2} \rightarrow
\rho^{\alpha,M}_{\sigma_1\sigma_2}$ and with extra terms coming from the regulators.  The Euclidean Pauli-Villars computation proceeds in precisely the same way and again agrees after analytic continuation.

Finally, we note that the analogous 1-loop correction
to the Wightman function $\langle \phi(X_1)\phi(X_2) \rangle$ of this
theory was recently considered by Krotov and Polyakov
(see \S 6 of \cite{Krotov:2010ma}; the same correlation function is
considered in \S 7, but with respect to a different state).
Our result for this correlation function is simply the right-hand side
of (\ref{eq:phi3InInFinal}) with the replacement
$\Delta_\mu(X_1,X_2) \to W_\mu(X_1,X_2)$. It is difficult to compare
these two results exactly because the result of  \cite{Krotov:2010ma}
has not been
renormalized (our renormalized result is presented in \cite{Marolf:2010zp}).
However, we can safely compare the behavior of the two results in the
infrared where the effect of renormalization is clear.
To compare with \cite{Krotov:2010ma} we set all masses to be equal.
Using techniques presented in \cite{Marolf:2010zp} we find the leading
behavior at large $|Z_{12}|\gg 1$ to be
\eqn{
  \label{eq:Polyakov}
  \left\langle \phi(X_1)\phi(X_2) \right\rangle
  % &=& g^2 \frac{\rho_{\s\s}^\a(\s)}{4(\s+\a)^2}
  % \left[ \partial_\mu \D_\mu(X_1,X_2) \right]_{\mu=\s}
  % \nn \\
  &=& \frac{g^2 \rho^\a_{\s\s}(\s) - \delta m^2 + m^2 \delta \phi}
  {16\pi^{\a+1}(\s+\a)^2}
  \nn \\ & &
  \times \big\{
  \Gamma(-\s)\Gamma(\s+\a)(-2 Z_{12})^\s \log Z_{12}
  \nn \\ & &
  + \Gamma(\s+2\a)\Gamma(-\s-\a)(-2 Z_{12})^{-(\s+2\a)} \log Z_{12}
  \big\}\left[1 + O\left(Z_{12}^{-1}\right) \right] . \nn \\
}
Here $\delta m^2$ and $\delta \phi$ are the real, divergent,
coefficients of the mass and field renormalization counterterms
which cancel the divergent terms in $\rho^\a_{\s\s}(\s)$.
We find the same asymptotic dependence on $Z_{12}$ as \cite{Krotov:2010ma}; in
particular, while the Wightman function of the free theory has
two asymptotic branches which decay like $Z_{12}^\s$ and $Z_{12}^{-(\s+2\a)}$,
the $O(g^2)$ correction has two asymptotic branches that each decay slower by
a multiplicative factor of $\log Z_{12}$.

The authors of \cite{Krotov:2010ma} interpret the appearance of the logarithm in
the asymptotic behavior (\ref{eq:Polyakov}) as an indication of an
``infrared correction'' to the correlator. Indeed, the logarithm
indicates that the 1-loop correction induces an $O(g^2)$ correction to
the mass parameter $\s$; as a result, the asymptotic expansion of the
correlator is altered in perturbation theory like
$(Z)^{\s+O(g^2)} = O(g^2) (Z_{12})^\s \log Z_{12} + O(g^4)$.
The $O(g^2)$ correction to $\s$ can be computed by performing the
sum over 1PI diagrams of the form of Figure~\ref{fig:1loop} (b).
This analysis was performed in detail \cite{Marolf:2010zp}. There it was
found that, at least for scalar fields with bare masses belonging to the
principal series of $SO(D,1)$, the $O(g^2)$ correction to $\s$ has a finite
negative real part (equivalently, the correction introduces a finite
negative imaginary part to the self-energy) which cannot be
removed with a local Hermitian counterterm. Thus the $O(g^2)$ correction
unambiguously \emph{increases}
the rate of decay of the 1PI-summed correlator so that this correlator
decays \emph{faster} than any free Wightman function. This agrees with
the analogous computation in flat-space where the 1PI-summed correlator
also enjoys an enhanced exponential rate of decay at large separations
\cite{Srednicki:2007qs}.

\section{Discussion}
\label{disc}

We have shown that Euclidean techniques and in-in perturbation theory on the Poincar\'e (a.k.a. cosmological) patch of de Sitter yield identical correlation functions for scalar field theories with positive masses.
This is in contrast with the situation for the in-in perturbation theory defined by global coordinates on de Sitter, where the corresponding factorization property fails \cite{Krotov:2010ma} and the in-in scheme contains infra-red divergences.  Our equivalence holds diagram by diagram and for any finite value of appropriate Pauli-Villars regulator masses.  It thus also holds for the fully renormalized diagrams.  While we focussed on non-derivative interactions, interactions involving derivatives can be handled in precisely the same way so long as additional Pauli-Villars subtractions are made as described in section \ref{prop}. We used a 3-step argument in the main text, though a more direct analytic continuation is described in appendix A.

As a check on the above arguments, we also explicitly calculated the one-loop propagator corrections due to both $\phi^3$ and $\phi^4$ interactions for all masses and in all dimensions in section \ref{numerics}. The Poincar\'e in-in and Euclidean calculations agreed precisely
\footnote{
We have also used a combination of analytic and numerical techniques to check agreement of Poincar\'e in-in and Euclidean correlators for the tree-level 3-point function for $D=4$ for $m^2 = 2$ (conformal coupling) and also for the one-loop correction to the 4-point function for $D=3$ and $m^2 = 3/4$ (also conformal coupling) evaluated at two pairs of coincident points.  Both of these diagrams are finite and require no regularization.  Our numerics indicate agreement to at least one part in $10^7$.  As these calculations do not yield significant insights, we have refrained from presenting the details.}.  We suspect that methods similar to those used in section \ref{numerics}, perhaps combined with Mellin-Barnes techniques as in \cite{Marolf:2010nz}, could be used to give a rather direct diagram-by-diagram proof of the equivalence of Euclidean and Poincar\'e in-in techniques, but we have not explored the details.

A number of points merit further discussion.  First,
some physicists have conjectured
that in-in calculations in the Poincar\'e patch lead to IR divergences, even for fields with $m^2 > 0$ due to contributions with vertices at large conformal time $\lambda$.  But there are clearly no such divergences in Euclidean signature.  So how can the two forms of perturbation theory agree diagram by diagram?  We believe that, if there are such divergences, they are better classified as ultra-violet (UV) divergences and are associated with the fact that the limit $\lambda \rightarrow \infty$ defines a null surface (the cosmological horizon) so that light-cone singularities can arise even at what appear to be large separations between points.

To a certain extent, the classification of these divergences as UV or IR in the cosmological patch may be a matter of semantics.  What is important is that any divergences may be cancelled using only local counter-terms.  This much is clear from our analysis:
We have seen that adding a Pauli-Villars regulator $M^2$ removes all divergences, and that the in-in and Euclidean calculations agree at all finite values of $M^2$.  This means that they have the same divergence structure in the limit $M^2 \rightarrow \infty$, and that divergences can be removed using the same sets of counter-terms.  But all divergences for massive theories on $S^D$ are clearly ultra-violet in nature and so are the same as on ${\mathbb R}^d$.  Local counter terms suffice to remove them.

Second, the reader will recall that the argument given in section~\ref{factorization} to show factorization (i.e., that the vertical sections of the contour at infinite past may be neglected) required the propagators to fall off at large timelike separations. Without such fall-off, the two formalisms should not agree.  Instead, analytic continuation of the Euclidean perturbation theory would give the terms of the in-in formalism, together with terms associated with integrals over some contour at infinity in the complex $t$-plane.   How then should we interpret this disagreement?  If the propagators do not fall off at large times, then integrals over the contour at infinity will generally diverge.  Thus, one would expect at most one formalism to give finite results.  Let us suppose that the Euclidean formalism is well-defined and finite.  If one can establish the appropriate positivity properties, then analytic continuation will define a good quantum state.  In this case, it would appear that any divergences of the in-in formalism are an unphysical artifact of this particular perturbative framework, and one might hope to better relate the two formalisms through an appropriate resummation of the divergent in-in formalism.

There is some potential for this scenario to hold in perturbative gravity.  For example, the tree-level three-point correlator constructed by Maldacena in \cite{Maldacena:2002vr} in the momentum space is IR divergent when inverse Fourier-transformed to position space.  On the other hand, the three-point function constructed  on $S^D$ using Euclidean propagators would have no IR divergences (see,
e.g.~\cite{Higuchi:2000ge,Higuchi:2001uv} for $D=4$).  It may therefore be interesting to re-examine the three-point function in Euclidean gravity.  However, we note that some physicists have raised objections to these propagators~\cite{Miao:2009hb,Miao:2010vs}. In addition, at least with generic gauge choices the Euclidean gravitational action is not bounded below (though see \cite{loll}) .  This means that one cannot rely on Osterwalder-Schr\"ader arguments~\cite{GJ} to guarantee that analytic continuation of the Euclidean correlators defines a positive-definite Hilbert space, and positivity would need to be verified.

The other possibility when propagators do not fall off is that both forms of perturbation theory are ill-defined.  This is the case for massless scalars on de Sitter.  But even here the divergences can be an artifact of the particular scheme for perturbation theory.  In \cite{Rajaraman:2010xd}, Rajaraman showed that, in the presence of a $\phi^4$ interaction with positive coefficient in the Hamiltonian, the Euclidean scheme can be resummed to give a new well-defined perturbation theory.  Since the Euclidean action is bounded below, the resulting Euclidean correlators will satisfy reflection-positivity and can be analytically continued to give a good state of the Lorentzian theory.

We close with a brief comment on other generalizations.  Recall that our first step was to verify that the usual connection between Euclidean methods and thermal in-in field theory on a static spacetime holds in the context of the de Sitter static patch.  It is clear that similar arguments will hold in the static regions of generic spacetimes with bifurcate Killing horizons, so long as the propagators again fall off sufficiently quickly at large separations.  For a particularly amusing application, consider the standard Minkowski space correlators (in the Minkowski vacuum) for which the usual perturbation theory integrates the vertices of Feynman diagrams over all of Minkowski space.  We now see that, so long as their arguments are taken to lie in, say, the right Rindler wedge, these correlators can in fact be computed using in-in perturbation theory in the Rindler wedge, and thus by integrating vertices of the in-in diagrams only over this Rindler wedge.  One would expect this fact to be well-known, but we have been unable to find any discussions in the literature.

\subsection*{Acknowledgements}
We thank Chris Fewster, Stefan Hollands, Ian McIntosh, Sasha Polyakov, Arvind Rajaraman, Albert Roura, Mark Srednicki and Takahiro Tanaka for useful discussions and correspondence.
DM and IM were supported in part by the US National Science Foundation under NSF grant PHY08-55415 and by funds from the University of California.
AH thanks the Astro-Particle Theory and Cosmology Group and Department of Applied Mathematics at University of Sheffield and Physics Department at UCSB for kind hospitality while part of this work was carried out.  His work at UCSB was supported by a Royal Society International Travel Grant.

\appendix

\section{Direct analytic continuation}

If an analytic function $f(z_1,\ldots,z_N)$ of $N$ variables is integrated over a real $N$-dimensional compact surface ${\mathcal S}$ with no boundary in $\mathbb{C}^N$ as
\begin{equation}
I = \int_{\mathcal S} f(z_1,\ldots,z_N)dz_1\wedge \cdots \wedge dz_N,
\end{equation}
we have $I=0$ as long as $f$ has no singularities on or inside ${\mathcal S}$ because the differential form $f(z_1,\ldots,z_N)dz_1\wedge \cdots \wedge dz_N$ is closed. This generalization of Cauchy's theorem can be used for the analytic continuation of correlators in the Euclidean formalism to those in the Poincar\'e in-in formalism.  In either formalism the integration is over a manifold of the form $M^n$, where $M$ is a real $D$-dimensional surface in complexified sphere, $\mathbb{S}^D$, and where $n$ is the number of internal vertices. We showed in section~\ref{factorization} that we can take $M=C_E\times S_h^{D-1}$ where $C_E$ is a contour similar to that shown in figure~\ref{fig:six} and where $S_h^{D-1}$ is a $D-1$ dimensional half-sphere in the Euclidean formalism.  On the other hand, we take $M=C_P\times \mathbb{R}^{D-1}$ where $C_P$ is a contour on the complex $\lambda$-plane with measure $d\lambda/\lambda^D$ (see figure~\ref{fig:nine}) in the Poincar\'e in-in formalism.
The generalized Cauchy's theorem together with the regularization of the propagator in section~\ref{regA} can be used to show that the amplitude, which is an integral over $M^n$, is analytically continued as an analytic function of the external points on $\mathbb{S}^D$ if $M$ can be deformed, with the external points moving and remaining on $M$, without letting it cross any singularities of the integrand\footnote{We expect that the integrals on all intermediate surfaces can be shown to converge by methods similar to those employed in sections~\ref{factorization} and \ref{analyticity}.}.  In this appendix we demonstrate that this deformation of the surface $M$ of integration from $S_1=C_E\times S_h^{D-1}$ (Euclidean formalism) to $S_2= C_P\times\mathbb{R}^{D-1}$ (Poinar\'e in-in formalism) can indeed be achieved.

We start with the surface $S_1$ for the Euclidean formalism.  It can be given in Poincar\'e coordinates as follows:
\begin{equation}
S_1 = \{(\Lambda e^{i\tau},\mathbf{X}e^{i\tau}): \tau \in (-\epsilon,\epsilon),\,\, \Lambda^2 - \|\mathbf{X}\|^2 = f(\tau)>0,\,\,\Lambda >0, \mathbf{X}\in \mathbb{R}^{D-1}\}, \label{app:S1}
\end{equation}
where $f(\tau)\to \infty$ as $\tau \to \pm\epsilon$.  This can be shown using the following relationship between the static and Poincar\'e coordinates, $(t,\theta,\hat{X})$ with $\hat{X}\cdot\hat{X} = 1$, and $(\lambda,\mathbf{x})$, respectively:
\begin{align}
e^{-2t} & = \lambda^2-\mathbf{x}\cdot\mathbf{x},\\
\hat{X^i}\sin\theta & = x^i/\lambda.
\end{align}
On the other hand the contour in figure \ref{fig:nine}, which is the $\lambda$-contour for the Poincar\'e in-in formalism before taking the limit ${\rm Im}\ \lambda\to 0$, corresponds to
\begin{equation}
S_2=\{(\left[f(\tau)\right]^{1/2}+ i\tau,\mathbf{X}): \tau \in(-\epsilon,\epsilon),\,\,\mathbf{X}\in\mathbb{R}^{D-1}\}.
\end{equation}

If the points $X_1=(\lambda_1,\mathbf{x}_1)$ and $X_2 = (\lambda_2,\mathbf{x}_2)$ are the arguments of a propagator, then (\ref{invP}) shows that it is singular if and only if
\begin{equation}
(X_1-X_2)^2 = -(\lambda_1-\lambda_2)^2 + (\mathbf{x}_1-\mathbf{x}_2)\cdot(\mathbf{x}_1-\mathbf{x}_2) = 0. \label{app:zero}
\end{equation}
It can readily be seen that this equation is not satisfied by any pair of distinct points on $S_1$ or $S_2$. Since the integrand is a product of propagators with arguments on the surface of integration, what we need to show is that there is a continuous deformation from $S_2$ to $S_1$ such that no intermediate surfaces contain two distinct points satisfying (\ref{app:zero})\footnote{We can show as in section~\ref{regA} that coincidence singularities do not spoil the analytic continuation argument here.}. We note that, if the vector ${\rm Im}\ X_1 - {\rm Im}\ X_2$ is timelike, then (\ref{app:zero}) does not hold.

First consider the following one-parameter family of surfaces:
\begin{equation}
S_{2,\gamma} = \{(\Lambda + i\tau,\mathbf{X}): \Lambda^2 - \gamma \|\mathbf{X}\|^2 = f(\tau)\},
\end{equation}
where $f(\tau)$ is the same positive function as in (\ref{app:S1}) and where $0 \leq \gamma \leq 1$.  Note that $S_{2,0} = S_2$. For any two points $X_j = (\Lambda_j+i\tau_j,\mathbf{X}_j)$, $j=1,2$, on $S_{2,\gamma}$,
we have
\begin{equation}
{\rm Im}\ X_1 - {\rm Im}\ X_2 = (\tau_1-\tau_2,\mathbf{0}),
\end{equation}
which is timelike if $\tau_1\neq \tau_2$.  If $\tau_1 = \tau_2$, then
$(X_1 - X_2)^2>0$
because $X_1=(\Lambda_1,\mathbf{X}_1)$ and $X_2=(\Lambda_2,\mathbf{X}_2)$ are both on the hyperboloid
$\Lambda^2 - \gamma\|\mathbf{X}\|^2 = f(\tau_1)$ with $0\leq \gamma\leq 1$.  Thus, the deformation of $S_2$ to $S_{2,1}$ leads to analytic continuation of the integral.

Next we consider the following two-parameter family of surfaces:
\begin{equation}
S_{(\alpha,\beta)} = \{((\Lambda + i\alpha \tau)e^{i\beta \tau}, \mathbf{X}e^{i\beta\tau}):
\Lambda^2 - \|\mathbf{X}\|^2 = f(\tau)\},
\end{equation}
where $0\leq \alpha,\beta \leq 1$.  We note that $S_{2,1}=S_{(1,0)}$ and $S_1 = S_{(0,1)}$.  Consider two points on $S_{(\alpha,\beta)}$:
\begin{align}
X_1 & = ((\Lambda_1+i\alpha\tau_1)e^{i\beta\tau_1},\mathbf{X}_1e^{i\beta\tau_1}),\\
X_2 & = ((\Lambda_2+i\alpha\tau_2)e^{i\beta\tau_2},\mathbf{X}_2e^{i\beta\tau_2}).
\end{align}
Define $\tilde{X}_j :=e^{-i\beta\tau_2}X_j$, $j=1,2$.  Since (\ref{app:zero}) is invariant under multiplication of $X_1$ and $X_2$ by a common factor, it is not satisfied if ${\rm Im}\ \tilde{X}_1 - {\rm Im}\ \tilde{X}_2$ is timelike. We find
\begin{equation}
{\rm Im}\ \tilde{X}_1 - {\rm Im}\ \tilde{X}_2
= (\Lambda_1\sin\beta(\tau_1-\tau_2)+ \alpha\tau_1\cos\beta(\tau_1-\tau_2) - \alpha \tau_2,
\mathbf{X}_1 \sin\beta(\tau_1-\tau_2)).
\end{equation}
If $\epsilon$ is sufficiently small --- recall $|\tau_1|,|\tau_2| < \epsilon$ --- then
this vector is timelike for $0\leq \alpha, \beta\leq 1$ provided that at least one of them is nonzero and that $\tau_1\neq \tau_2$.
If $\tau_1=\tau_2$, then we have
$(\tilde{X}_1-\tilde{X}_2)^2>0$ because $\tilde{X}_1=(\Lambda_1,\mathbf{X}_1)$ and $\tilde{X}_2=(\Lambda_2,\mathbf{X}_2)$ are both on the hyperboloid
$\Lambda^2 - \|\mathbf{X}\|^2 = f(\tau_1)$.
Thus, the deformation of $S_{2,1}$ to $S_1$ gives analytic continuation of the integral if $\epsilon$ is sufficiently small, and, hence, so does the deformation of $S_2$ to $S_1$.  When combined with the various convergence and fall-off arguments from the main text, this result implies that the correlators computed using the contour of figure~\ref{fig:nine} in the Poincar\'e in-in formalism is equal to the corresponding analytic continuation of the Euclidean correlators.

%%%%%%%%%%%%%%%%%%%%%%%%%%%%%%%%%%%%%%%%%%%%%%%%%%%%%%%%%%%%%%%%%

%%%%%%%%%%%%%%%%%%%%%%%%%%%%%%%%%%%%%%%%%%%%%%%%%%%%%%%%%%%%%%%%%

%%%%%%%%%%%%%%%%%%%%%%%%%%%%%%%%%%%%%%%%%%%%%%%%%%%%%%%%%%%%%%%%%

\begin{thebibliography}{99}
%%%%%%%%%%%%%%%%%%%%%%%%%%%%%%%%%%%%%%%%%%%%%%%%%%%%%%%%%%%%%%%%%

%\cite{Allen:1985ux}
\bibitem{Allen:1985ux}
  B.~Allen,
  ``Vacuum States In De Sitter Space,''
  Phys.\ Rev.\  D {\bf 32}, 3136 (1985).
  %%CITATION = PHRVA,D32,3136;%%

\bibitem{AAS}
 A.~A.~Starobinsky,
  ``Spectrum of relict gravitational radiation and the early state of the
  universe,''
  JETP Lett.\  {\bf 30}, 682 (1979)
  [Pisma Zh.\ Eksp.\ Teor.\ Fiz.\  {\bf 30}, 719 (1979)].
  %%CITATION = ZFPRA,30,719;%%

\bibitem{EM}
  E.~Mottola,
  ``Particle Creation In De Sitter Space,''
  Phys.\ Rev.\  D {\bf 31}, 754 (1985);
  %%CITATION = PHRVA,D31,754;%%
  E.~Mottola, ``Fluctuation - dissipiation theorem in general relativity and the
  cosmological constant,'' {\it Physical Origins of Time Asymmetry} (Cambridge, Cambridge Univ. Press 1993) ed by J. J. Halliwell et al, pp. 504-515;
  I.~Antoniadis, P.~O.~Mazur and E.~Mottola,
  ``Cosmological dark energy: Prospects for a dynamical theory,''
  New J.\ Phys.\  {\bf 9}, 11 (2007)
  [arXiv:gr-qc/0612068];
  %%CITATION = NJOPF,9,11;%%
 E.~Mottola,
  ``New Horizons in Gravity: The Trace Anomaly, Dark Energy and Condensate
  Stars,''
  arXiv:1008.5006 [gr-qc].
  %%CITATION = ARXIV:1008.5006;%%

\bibitem{Hu:1985uy}
  B.~L.~Hu and D.~J.~O'Connor,
  ``Infrared Behavior And Finite Size Effects In Inflationary Cosmology,''
  Phys.\ Rev.\ Lett.\  {\bf 56}, 1613 (1986).
  %%CITATION = PRLTA,56,1613;%%

\bibitem{Hu:1986cv}
  B.~L.~Hu and D.~J.~O'Connor,
  ``Symmetry Behavior in Curved Space-Time: Finite Size Effect and Dimensional Reduction,''
  Phys.\ Rev.\  D {\bf 36}, 1701 (1987).
  %%CITATION = PHRVA,D36,1701;%%


\bibitem{TW}  N.~C.~Tsamis and R.~P.~Woodard,
  ``Relaxing The Cosmological Constant,''
  Phys.\ Lett.\ B {\bf 301}, 351 (1993);
  %%CITATION = PHLTA,B301,351;%%
  N.~C.~Tsamis and R.~P.~Woodard,
  ``Strong infrared effects in quantum gravity,''
  Annals Phys.\ {\bf 238}, 1 (1995);
  %%CITATION = APNYA,238,1;%%
  N.~C.~Tsamis and R.~P.~Woodard,
  ``Quantum Gravity Slows Inflation,''
  Nucl.\ Phys.\  B {\bf 474}, 235 (1996)
  [arXiv:hep-ph/9602315];
  %%CITATION = NUPHA,B474,235;
  ``The quantum gravitational back-reaction on inflation,''
  Annals Phys.\  {\bf 253}, 1 (1997)
  [arXiv:hep-ph/9602316];
  %%CITATION = APNYA,253,1;%%  N.~C.~Tsamis and R.~P.~Woodard,
  ``Stochastic quantum gravitational inflation,''
  Nucl.\ Phys.\  B {\bf 724}, 295 (2005)
  [arXiv:gr-qc/0505115].
  %%CITATION = NUPHA,B724,295;%%

%\bibitem{Higuchi:2002sc}
 %A.~Higuchi and R.~H.~Weeks,
  %``The physical graviton two-point function in de Sitter spacetime with S(3)
  %spatial sections,''
  %Class.\ Quant.\ Grav.\  {\bf 20}, 3005 (2003)
  %[arXiv:gr-qc/0212031].
  %%%CITATION = CQGRD,20,3005;%%



\bibitem{polyakov1}
  A.~M.~Polyakov,
  ``De Sitter Space and Eternity,''
  Nucl.\ Phys.\  B {\bf 797}, 199 (2008)
  [arXiv:0709.2899 [hep-th]].
  %%CITATION = NUPHA,B797,199;%%

\bibitem{PerezNadal:2008ju}
  G.~Perez-Nadal, A.~Roura, E.~Verdaguer,
  ``Backreaction from non-conformal quantum fields in de Sitter spacetime,''
  Class.\ Quant.\ Grav.\  {\bf 25}, 154013 (2008).
  [arXiv:0806.2634 [gr-qc]].


\bibitem{Faizal:2008ns}
  M.~Faizal and A.~Higuchi,
  ``On the FP-ghost propagators for Yang-Mills theories and perturbative
  quantum gravity in the covariant gauge in de Sitter spacetime,''
  Phys.\ Rev.\  D {\bf 78}, 067502 (2008)
  [arXiv:0806.3735 [gr-qc]].
  %%CITATION = PHRVA,D78,067502;%%

\bibitem{Akhmedov:2008pu}
  E.~T.~Akhmedov, P.~V.~Buividovich,
  ``Interacting Field Theories in de Sitter Space are Non-Unitary,''
  Phys.\ Rev.\  {\bf D78}, 104005 (2008).
  [arXiv:0808.4106 [hep-th]].


\bibitem{Higuchi:2009zza}
  A.~Higuchi,
  ``Decay of the free-theory vacuum of scalar field theory in de Sitter
  spacetime in the interaction picture,''
  Class.\ Quant.\ Grav.\  {\bf 26}, 072001 (2009) [arXiv:0809.1255 [gr-qc]].
  %%CITATION = CQGRD,26,072001;%%


\bibitem{Higuchi:2009ew}
  A.~Higuchi and Y.~C.~Lee,
  ``A conformally-coupled massive scalar field in de Sitter expanding universe
  with the mass term treated as a perturbation,''
  arXiv:0903.3881 [gr-qc].
  %%CITATION = ARXIV:0903.3881;%%

\bibitem{Akhmedov:2009ta}
  E.~T.~Akhmedov,
  ``Real or Imaginary? (On pair creation in de Sitter space),''
  [arXiv:0909.3722 [hep-th]].




\bibitem{Polyakov:2009nq}
  A.~M.~Polyakov,
  ``Decay of Vacuum Energy,''
  arXiv:0912.5503 [hep-th].
  %%CITATION = ARXIV:0912.5503;%%

\bibitem{Burgess:2010dd}
  C.~P.~Burgess, R.~Holman, L.~Leblond and S.~Shandera,
  ``Breakdown of Semiclassical Methods in de Sitter Space,''
  arXiv:1005.3551 [hep-th].
  %%CITATION = ARXIV:1005.3551;%%



%\cite{Giddings:2010nc}
\bibitem{Giddings:2010nc}
  S.~B.~Giddings and M.~S.~Sloth,
  ``Semiclassical relations and IR effects in de Sitter and slow-roll
  space-times,''
  arXiv:1005.1056 [hep-th].
  %%CITATION = ARXIV:1005.1056;%%

  %\cite{Krotov:2010ma}
\bibitem{Krotov:2010ma}
  D.~Krotov, A.~M.~Polyakov,
``Infrared Sensitivity of Unstable Vacua,''
[arXiv:1012.2107 [hep-th]].


\bibitem{Hajicek}
P.~H\'aj\'{\i}\v{c}ek, ``A new generating functional for expectation values of field operators,'' Bern preprint, 1978 (unpublished).

\bibitem{Kay80}
B.~S.~Kay, ``Linear spin-zero quantum fields in external gravitational and scalar fields. II. Covarivant perturbation theory'', Commun.\ Math.\ Phys.\ {\bf 71}, 29 (1980).

%\cite{Jordan:1986ug}
\bibitem{Jordan:1986ug}
  R.~D.~Jordan,
  ``Effective Field Equations for Expectation Values,''
  Phys.\ Rev.\  {\bf D33}, 444-454 (1986).

%\cite{Calzetta:1986ey}
\bibitem{Calzetta:1986ey}
  E.~Calzetta, B.~L.~Hu,
  %``Closed Time Path Functional Formalism in Curved Space-Time: Application to Cosmological Back Reaction Problems,''
  Phys.\ Rev.\  {\bf D35}, 495 (1987).




\bibitem{Hartle:1976tp}
  J.~B.~Hartle and S.~W.~Hawking,
  ``Path Integral Derivation Of Black Hole Radiance,''
  Phys.\ Rev.\  D {\bf 13}, 2188 (1976).
  %%CITATION = PHRVA,D13,2188;%%


\bibitem{Marolf:2010zp}
  D.~Marolf and I.~A.~Morrison,
  ``The IR stability of de Sitter: Loop corrections to scalar propagators,''
 Phys.\ Rev.\ {\bf D82}, 105032 (2010) [arXiv:1006.0035 [gr-qc]].
  %%CITATION = ARXIV:1006.0035;%%

%\cite{Marolf:2010nz}
\bibitem{Marolf:2010nz}
  D.~Marolf, I.~A.~Morrison,
  ``The IR stability of de Sitter QFT: results at all orders,''
arXiv:1010.5327 [gr-qc].


\bibitem{Hollands:2010pr}
  S.~Hollands,
  ``Correlators, Feynman diagrams, and quantum no-hair in deSitter spacetime,''
  [arXiv:1010.5367 [gr-qc]].

\bibitem{Rajaraman:2010xd}
  A.~Rajaraman,
  ``On the proper treatment of massless fields in Euclidean de Sitter space,''
  arXiv:1008.1271 [hep-th].
  %%CITATION = ARXIV:1008.1271;%%

\bibitem{Schlingemann:1999mk}
  D.~Schlingemann,
  ``Euclidean field theory on a sphere,''
  arXiv:hep-th/9912235.
  %%CITATION = HEP-TH/9912235;%%

\bibitem{GJ}
J. Glimm and A. Jaffe, {\it Quantum Physics} (Springer-Verlag, New York, 1987), sections 6.1 and 10.4.

%\cite{Gibbons:1976pt}
\bibitem{Gibbons:1976pt}
  G.~W.~Gibbons, M.~J.~Perry,
  ``Black Holes and Thermal Green's Functions,''
  Proc.\ Roy.\ Soc.\ Lond.\  {\bf A358}, 467-494 (1978).

%\cite{Gibbons:1977mu}
\bibitem{Gibbons:1977mu}
  G.~W.~Gibbons, S.~W.~Hawking,
  ``Cosmological Event Horizons, Thermodynamics, and Particle Creation,''
  Phys.\ Rev.\  {\bf D15}, 2738-2751 (1977).

%\cite{Schwinger:1960qe}
\bibitem{Schwinger:1960qe}
  J.~S.~Schwinger,
  ``Brownian motion of a quantum oscillator,''
  J.\ Math.\ Phys.\  {\bf 2}, 407-432 (1961).

%\cite{Keldysh:1964ud}
\bibitem{Keldysh:1964ud}
  L.~V.~Keldysh,
  ``Diagram technique for nonequilibrium processes,''
  Zh.\ Eksp.\ Teor.\ Fiz.\  {\bf 47}, 1515-1527 (1964).

\bibitem{Landsman:1986uw}
  N.~P.~Landsman and C.~G.~van Weert,
  ``Real and Imaginary Time Field Theory at Finite Temperature and Density,''
  Phys.\ Rept.\  {\bf 145}, 141 (1987).
  %%CITATION = PRPLC,145,141;%%

%\cite{Bunch:1978yq}
\bibitem{Bunch:1978yq}
  T.~S.~Bunch, P.~C.~W.~Davies,
  ``Quantum Field Theory in de Sitter Space: Renormalization by Point Splitting,''
  Proc.\ Roy.\ Soc.\ Lond.\  {\bf A360}, 117-134 (1978).

\bibitem{Allen:1985wd}
  B.~Allen and T.~Jacobson,
  ``Vector Two Point Functions In Maximally Symmetric Spaces,''
  Commun.\ Math.\ Phys.\  {\bf 103}, 669 (1986).
  %%CITATION = CMPHA,103,669;%%


\bibitem{Vilenkin91}
  N.~Ya.~Vilenken, and A.~U.~Klimyk,
  ``Representations of Lie Groups and Special Functions,'' vols 1-3.
  (Dordrecht: Kluwer Acad. Publ. 1991-1993).

\bibitem{Camporesi:1992wn}
  R.~Camporesi and A.~Higuchi,
  ``Stress Energy Tensors In Anti-De Sitter Space-Time,''
  Phys.\ Rev.\  D {\bf 45}, 3591 (1992).
  %%CITATION = PHRVA,D45,3591;%%

%\cite{Streater:1989vi}
\bibitem{Streater:1989vi}
  R.~F.~Streater and A.~S.~Wightman,
  ``PCT, spin and statistics, and all that,''
  Redwood City, USA: Addison-Wesley (1989) 207 p. (Advanced book classics).


\bibitem{Collins:1984xc}
  J.~C.~Collins,
  ``Renormalization. An Introduction To Renormalization, The Renormalization
  Group, And The Operator Product Expansion,''
%\href{http://www.slac.stanford.edu/spires/find/hep/www?irn=1341391}{SPIRES entry}
{\it  Cambridge, Uk: Univ. Pr. (1984) 380p}

%\cite{Srednicki:2007qs}
\bibitem{Srednicki:2007qs}
  M.~Srednicki,
  ``Quantum field theory,''
%\href{http://www.slac.stanford.edu/spires/find/hep/www?irn=7209290}{SPIRES entry}
{\it  Cambridge, UK: Univ. Pr. (2007) 641 p}



\bibitem{Maldacena:2002vr}
  J.~M.~Maldacena,
  ``Non-Gaussian features of primordial fluctuations in single field
  inflationary models,''
  JHEP {\bf 0305}, 013 (2003)
  [arXiv:astro-ph/0210603].
  %%CITATION = JHEPA,0305,013;%%

\bibitem{Higuchi:2000ge}
 A.~Higuchi and S.~S.~Kouris,
  ``On the scalar sector of the covariant graviton two-point function in  de
  Sitter spacetime,''
  Class.\ Quant.\ Grav.\  {\bf 18}, 2933 (2001)
  [arXiv:gr-qc/0011062].
  %%CITATION = CQGRD,18,2933;%%

  \bibitem{Higuchi:2001uv}
  A.~Higuchi and S.~S.~Kouris,
  ``The covariant graviton propagator in de Sitter spacetime,''
  Class.\ Quant.\ Grav.\  {\bf 18}, 4317 (2001)
  [arXiv:gr-qc/0107036].
  %%CITATION = CQGRD,18,4317;%%

 \bibitem{Miao:2009hb}
  S.~P.~Miao, N.~C.~Tsamis and R.~P.~Woodard,
  ``Transforming to Lorentz Gauge on de Sitter,''
  J.\ Math.\ Phys.\  {\bf 50}, 122502 (2009)
  [arXiv:0907.4930 [gr-qc]].
  %%CITATION = JMAPA,50,122502;%%


 \bibitem{Miao:2010vs}
  S.~P.~Miao, N.~C.~Tsamis and R.~P.~Woodard,
  ``De Sitter Breaking through Infrared Divergences,''
  J.\ Math.\ Phys.\  {\bf 51}, 072503 (2010)
  [arXiv:1002.4037 [gr-qc]].
  %%CITATION = JMAPA,51,072503;%%

\bibitem{loll}
  A.~Dasgupta, R.~Loll,
  ``A Proper time cure for the conformal sickness in quantum gravity,''
  Nucl.\ Phys.\  {\bf B606}, 357-379 (2001).
  [hep-th/0103186].



%%%%%%%%%%%%%%%%%%%%%%%%%%%%%%%%%%%%%%%%%%%%%%%%%%%%%%%%%%%%%%%%%
\end{thebibliography}
\end{document}